\documentclass[12pt]{iopart}
\usepackage{color,xcolor}
\usepackage{float}
\usepackage{ulem}
\usepackage{graphicx}
\usepackage{subfigure}
\usepackage{longtable}
\usepackage{etoolbox}
\usepackage{gensymb}
\usepackage{fancyhdr}
\usepackage{hyperref}
\hypersetup{
    colorlinks=true,
    linkcolor=blue,
    filecolor=magenta, 
    citecolor = blue,
}
\newcommand{\Tilde}[1]{\widetilde{#1}}

\begin{document}

\title[A linear model of synergetic current drive with LHW and ECW]{A linear model of synergetic current drive with lower-hybrid wave and electron cyclotron wave}

\author{J.N. Chen$^{1,2,3,4}$, Y.M Zhang$^{4, 5}$, S.Y. Chen$^{4,5\dagger}$, M.L. Mou$^{4,5}$, C.J. Tang$^{4,5}$}

\address{$^{1}$ School of Fundamental Physics and Mathematical Sciences, Hangzhou Institute for Advanced Study, UCAS, Hangzhou 310024, China}
\address{$^{2}$ Institute of Theoretical Physics, Chinese Academy of Science, Beijing 100190, China}
\address{$^{3}$ University of Chinese Academy of Sciences, Beijing 100049, China}
\address{$^{4}$ College of Physics, Sichuan University, Chengdu 610065, China} 
\address{$^{5}$ Key Laboratory of High Energy Density Physics and Technology of Ministry of Education, Sichuan University, Chengdu 610064, China}

\ead{$^{\dagger}$sychen531@163.com}
\vspace{10pt}

\begin{indented}
\item[]August 2025
\end{indented}

\begin{abstract}

A linear model of synergetic current drive (SCD) with lower-hybrid wave (LHW) and electron cyclotron wave (ECW) is developed to quickly calculate the quantitative SCD efficiency and reveal the conditions for the occurrence of the synergy effect. In this model, the response function dominated by collisions in the presence of LHW is derived from the adjoint equations by using perturbation and Green-Function techniques, where the relativistic and trapping effects are considered. The SCD results compared to LinLiu's ECW current drive (ECCD) efficiency show two features of the synergy effect, one is that it is inclined to occur at smaller $y = 2\omega_{c}/\omega$ with the fixed ECW parallel refractive index $n_{\parallel}$, and the other is that the threshold values of $y$, at which the synergy effect becomes sufficiently significant, shifts towards higher values with a decreasing $n_{\parallel}$; the SCD results without the trapping effect and the response function without relativistic effect have been provided. The quasilinear simulation on ECCD and SCD efficiency with a two-dimensional Fokker-Planck code is consistent with the results of the linear model in trends. Given considerable parameters improvement nowadays, the SCD efficiencies with parameters of current mainstream reactors are calculated in this work. We provide the current profile of ECCD and SCD by combining with GENRAY, showing the potential application of integrating SCD into ray-tracing code. Criteria for the occurrence and the sufficient significance of the synergy effect are suggested, which indicate that the synergy effect is dependent on the power factor that quantifies the degree of the overlap of the two waves’ quasilinear domains, the LHW power, and synergy electrons.

\end{abstract}
\vspace{2pc}
\noindent{\it Keywords}: current drive, the synergy effect, lower-hybrid wave, electron cyclotron wave. \\
\noindent{\submitto{\NF}}

% \tableofcontents

\section{Introduction}

In Tokamak, the injection of radio frequency waves (RF) for current drive (CD)\cite{Fish1987_RevModPhys.59.175} to realize active control of plasma current profile can not only stabilize MHD but also is crucial to achieving steady-state long-pulse operation in the future reactor. The principle of RF CD is to use an external device to inject asymmetric momentum into plasmas or establish asymmetric resistivity. The typical applications for experiments corresponding to different mechanisms are lower-hybrid wave current drive (LHCD) and electron cyclotron wave current drive (ECCD). Both of them transfer the energy of the wave to resonant particles, and then a net particle flow(i.e., current) is driven.  LHCD has high CD efficiency, but its current profile is not easy to control, and ECCD has high locality and positive maneuverability, but CD efficiency is small. Based on their complementary properties, the idea of synergetic current drive~(SCD) has been proposed~\cite{Fidone198410.1063/1.864528} to optimize the driven current. Subsequently, both the simulations using Fokker-Planck code~\cite{Fidone_1987,SHOUCRI1989253}, and the model combining the kinetic process~\cite{Dumont2000combined} manifest that the main feature of SCD is that the current driven by lower-hybrid waves (LHW) and electron cyclotron waves (ECW) simultaneously $I_{LH+EC}$ can be larger than the sum of currents driven individually $I_{LH}+I_{EC}$ in the same plasma conditions, hereafter called the synergy effect. SCD was first achieved in stationary conditions on the Tore Supra~\cite{PhysRevLett.93.255002}, the core experimental phenomenon is the improvement(up to a factor of $\sim 4$) of ECCD efficiency sustained by LHCD, but the LHCD efficiency is weakly increased in presence of ECW, and this experiment demonstrated that the synergy effect is considered as the most reasonable interpretation of ECCD efficiency improvement; the most promising application of SCD is off-axis current profile control.

Numerical simulations provide more valuable insights into SCD, despite their time-consuming nature. The mechanism of the synergy effect can be simply explicated in velocity space:  ECWs push resonant electrons to higher $u_{\perp}$, thus allowing some electrons to reach the low $u_{\parallel}$ and high $u_{\perp}$ limit of the LHWs interaction region via pitch-angle scattering, these additional electrons can be further accelerated by LHWs, which are the origin of the synergy effect~\cite{PhysRevLett.93.255002, Dumont2004theory}. However, the synergy effect is influenced by the radial location $\rho_{EC}$ of maximum ECW power deposition~\cite{PhysRevLett.93.255002} and the resonance relationship between LHW and ECW in velocity space\cite{Dumont2004theory}. When there is an overlap of power deposition of LHW and ECW in radial location, as well as an overlap of the quasilinear domains in velocity space, the synergy effect occurs. The dependence of this effect on the frequency $\omega$ and parallel refractive index $n_{\parallel}$ of ECW has been investigated~\cite{Huang_22252162}. Additionally, other kinetic simulations have shown a negative effect\cite{negative_Chen_2012} when the ECW power greatly exceeds the LHW power, and a new synergy mechanism\cite{OKCD_Zheng_2019} between ECCD based on the Ohkawa effect and LHCD has been observed even in the absence of overlapping quasilinear domains.

There is less work investigating SCD theoretically. The understanding of the synergy effect remains qualitative in terms of theory, although various properties of SCD have been revealed in simulations, and the quantitative current can be calculated. The pioneering work~\cite{Dumont2004theory} from the theoretical analysis has qualitatively proved the existence of the synergy effect by using the perturbation technique, while this work doesn't take the trapped electrons and the relativistic effect into account since only the simplest Fisch-Boozer response function is applied to calculate the perturbative response function induced by LHW and the CD efficiency. However, ECCD~\cite{Cohen1987,LinLiu2003} and LHCD\cite{Fish1987_RevModPhys.59.175, D.W.Ignat_1994} both have been investigated quantitatively from a theoretical perspective and integrated into ray-tracing codes such as TORAY-GA\cite{toray_matsuda1989ray} and GENRAY\cite{genray_smirnov1994general}, which benefit fusion plasma simulation a lot. Therefore, developing a linear model of SCD can not only rapidly calculate SCD efficiency quantitatively and deepen the understanding of the synergy effect, but also have enormous application potential that integrates the linear model into ray-tracing codes. 

We develop a linear model of SCD in terms of theory in this paper. For the aim to establish it, the adjoint method, the technique of perturbation, the Green-Function solution, and Legendre expansion in pitch-angle $\xi = v_{\parallel}/v$ \cite{Antonsen198210.1063/1.863906, Fish1987_RevModPhys.59.175, Dumont2004theory, LinLiu2003},  have been applied to solve the response function in presence of LHW at low-collisionality regime\cite{Fisch1981_PhysRevA.24.3245} and calculate SCD efficiency. The relativistic and the trapped electrons are taken into account in our model; they haven't been considered in previous theoretical works simultaneously. The magnetic well is approximated as a square well for convenience. Firstly, we calculate the SCD efficiency using the same plasma parameters as those in the widely used ECCD model~\cite{LinLiu2003}, where the applied extraordinary mode~(X-mode) of ECWs, for benchmarking this linear model with it. Then, we show the derivations and efficiency of SCD without trapped electrons. Due to significant improvements in current reactors, we next provide the SCD results utilizing the parameters of ITRE, DEMO1, and CFETR, where the ordinary mode~(O-mode) of ECWs is a primary resonant mechanism. The current profiles of ECCD and SCD, combined with GENRAY~\cite{genray_smirnov1994general}, are provided, showing the potential application of integrating SCD into ray-tracing code.

The organization of the paper is as follows. Section \ref{sec: establish the model} focuses on establishing the linear model. Section \ref{sec: results} presents the main results and discussion based on the model with specific parameters. Section \ref{sec: criteria} provides the criteria for measuring the synergy effect. Finally, a summary is in Section \ref{sec: summary}. \ref{apd: Details of Lin-Liu's response function}  presents the details of the zero-order response function $\tilde{\chi}_{0}$. In  \ref{apd: The approximation of Fu}, we evaluate the effectiveness of the approximation that takes $F(u)$ as a simple function with fixed order and coefficient in numerical calculation. \ref{apd: The interpolation formulation of E_lambda} presents the interpolation formula for the angular part of the perturbation response function.

\section{The linear model of SCD} \label{sec: establish the model}

After RF wave injection into plasma, the motion of electrons is characterized by the relativistic dynamics in the following derivation. Consider $\mathbf{u} = \mathbf{p}/m = \gamma \mathbf{v} $ with $\gamma = \sqrt{1 + (u/c)^2}$ denoting the momentum per unit mass, and the kinetic particle energy is $w = (\gamma - 1)mc^2$. On the premise of low RF power density, the electron distribution function in the presence of RF satisfies 
\begin{equation}
	f = f_{M} + \delta f ,
\end{equation}
where $f_{M}$ is the Maxwellian distribution, $\delta f$ is the perturbed distribution function caused by external energy injection, both of them are considered as a function of kinetic energy $w$, magnetic moment  $\mu = m u_{\perp}^{2}/2B$, and poloidal angle of a given flux surface $\theta_{p}$. The premise also implies that the interaction between waves and electrons can be portrayed by the RF-induced quasilinear diffusion operator in velocity space $S_{\mathrm{rf}}(f_{M})$. Ignoring the small cross-effect drifts and finite banana width, the linearized Fokker-Planck equation, which describes the dynamic behavior of the perturbed distribution $\delta f$, can be written as
\begin{equation}
	v_{\parallel} \hat{\mathbf{b}} \cdot \nabla \delta f - \hat{O}\delta f = S_{\mathrm{rf}}\left(f_{M}\right),
\end{equation}
where $\hat{O}$ denotes a linearized operator and $\hat{\mathbf{b}}$ represents the direction of local magnetic field~($\hat{b}=\mathbf{B}/B$); the RF-driven current is parallel to the local magnetic field~($\mathbf{j_{rf}} =  j_{\parallel} \vec{\mathbf{b}}$). The response function $\chi$ associated with the driven current that is caused by the perturbed function satisfies
\begin{equation} \label{general equation satisfy the response function}
	-v_{\parallel} \hat{\mathbf{b}} \cdot \nabla \chi- \hat{O}^{\dagger}\chi=\frac{v_{\parallel} B}{\left\langle B^{2}\right\rangle},
\end{equation}
where $\hat{O}^\dagger$ is the adjoint operator of $\hat{O}$ :
\begin{equation}
 	\int d\mathbf{u} ( \hat{O}^{\dagger} h(\mathbf{u}) ) g(\mathbf{u}) = \int d\mathbf{u} h(\mathbf{u}) \hat{O} g(\mathbf{u}).
\end{equation} 
The CD efficiency and the dimensionless CD efficiency~\cite{C.C.Petty_2002} are
\begin{equation} \label{eq: general CD efficiency}
	\zeta^{\star}  =  \frac{e^{3}}{\varepsilon_{0}^{2}} \frac{n_{e}}{T_{e}} \frac{\left\langle j_{\parallel}\right\rangle}{2 \pi Q} 
	= -\frac{4}{\ln \Lambda}\left\langle\frac{B}{B_{m }}\right\rangle \frac{\left\langle\int d \Gamma \tilde{\chi} S_{\mathrm{rf}}\left(f_{M}\right)\right\rangle}{\left\langle\int d \Gamma\left(w / m v_{e}^{2}\right) S_{\mathrm{rf}}\left(f_{M}\right)\right\rangle} ,
\end{equation}
\begin{equation}
	\tilde{\zeta}= \frac{e^3}{\varepsilon_{0}^{2}} \frac{n_{e}}{T_{e}} \frac{R_{p} I }{P} =\frac{R_{p}}{\langle B \rangle} \langle \frac{B_{\phi}}{R} \rangle \zeta^{\star}.
\end{equation}
with
\begin{equation}
	\left\langle j_{\parallel}\right\rangle =   -e  \langle\int d \Gamma \delta f v_{\parallel} \rangle 
	=  -e  \langle B \int d \Gamma \chi S_{\mathrm{rf}}\left(f_{M}\right)\rangle ,
\end{equation}
\begin{equation}
	Q=\left\langle\int d \Gamma w S_{\mathrm{rf}}\left(f_{M}\right)\right\rangle ,
\end{equation}
\begin{equation}
	\langle\cdots\rangle=\frac{\oint \frac{d \ell_{p}}{B_{p}} \ldots}{\oint \frac{d \ell_{p}}{B_{p}}} ,
\end{equation}
\begin{equation} \label{the relation between dimensionless response function and response function}
	\tilde{\chi}=\nu_{e 0}\left(B_{m } / v_{e}\right) \chi .
\end{equation}
Here $\varepsilon_{0}$ is the permittivity of free space, $\langle j_{\parallel} \rangle$ is the wave-driven current, $Q$ is the absorbed wave power density, $B_{m}$ is the maximum of $B$ at the flux surface, $d\Gamma$ is the volume element in velocity space, $\langle ... \rangle$ represents the flux-surface average, $dl_{p}$ is the line element along the poloidal circumference, $B_{p}$ is the poloidal magnetic field, $\tilde{\chi}$ is the dimensionless response function of $\chi$. $n_{e}$, $T_{e}$, $v_{e}=\sqrt{2T_{e}/m}$, $\ln \Lambda$, and $\nu_{e 0}=(e^{4}n_{e} ln\Lambda )/(4 \pi \epsilon_{0}^{2} m^{2} v_{e}^{3})$ are electron density, electron temperature, electron thermal velocity, Coulomb logarithm, and collision frequency, respectively.

The theoretical formulation of RF CD in the above derivations is universal as long as the wave injection does not significantly make the electron distribution deviate from the Maxwellian distribution, which is consistent with the premise of low RF power density. In ECCD case~\cite{LinLiu2003}, $\hat{O}$ and $S_{rf}(f_{M})$, respectively,  are replaced by the linearized collision operator $\hat{C}$ and the operator indicating an interaction between ECW and electrons $S_{ECW}(f_{M})$. Here $\hat{C}$ in Fisch's relativistic high-velocity model~\cite{Fish1987_RevModPhys.59.175} can be written as  
\begin{equation} \label{eq: the collision operator}
	\hat{C} f=\left[\nu_{e i}(u)+\nu_{D}(u)\right] \hat{L} f+\frac{1}{u^{2}} \frac{\partial}{\partial u} u^{2} \lambda_{s}(u) f ,
\end{equation}
where $\hat{L}$ denotes the pitch-angle scattering operator
\begin{equation}
	\hat{L} = \frac{1}{2} \frac{\partial}{\partial \xi}\left(1-\xi^{2}\right) \frac{\partial}{\partial \xi} ,
\end{equation}
with $	\xi = \frac{v_{\parallel}}{v}$, and the pitch-angle scattering rates because of electron-ion and electron-electron collisions are $\nu_{e i}(u)=Z_{\mathrm{eff}} \nu_{e 0} \gamma\left( u_{e}/u \right)^{3}$ and $\nu_{D}(u)=\nu_{e 0} \gamma\left(u_{e}/u \right)^{3} $ in which $Z_{eff}$ is the effective charge and $u_{e} = v_{e} = \sqrt{2T_{e}/m}$. The last term of Eq.(\ref{eq: the collision operator}) describes the slowing-down effect because of electron-electron collisions and the rate is $\lambda_{s}(u)=\nu_{e 0} u_{e} \gamma^{2}\left(\frac{u_{e}}{u}\right)^{2} $.
Additionally, the adjoint operator of $\hat{C}$ is given by
\begin{equation}
	\hat{C}^{\dagger} g=\left[\nu_{e i}(u)+\nu_{D}(u)\right] \hat{L} g-\lambda_{s}(u) 	\frac{\partial}{\partial u} g .
\end{equation}

Based on the experimental phenomenon of the synergy effect, which is the improvement of ECCD efficiency in the presence of LHW, following R.Dumont~\cite{Dumont2004theory}, we consider the electron relaxation is dominated by collisions and diffusions caused by the LHW power in a given velocity space together, where collisions play the leading role, such that operator $\hat{O}$ will be regarded as $\hat{C} + \hat{D_{LH}}$, and $S_{rf}(f_M)$ is the same as the ECCD case which describes the interaction between ECW and electrons. It implies that ECW will act on the electron distribution altered by the injection of the LHW power $f = f_{M}(1+\phi)$, in which $f_{M}\phi$ represents the modification caused by the LHW effect and is also a perturbation quantity so that the SCD efficiency calculated in this situation by using $ S_{\mathrm{rf}}(f) \simeq S_{\mathrm{rf}}({f_{M}})$ is still effective. Here $\hat{D}_{LH}$ is a linearized operator describing diffusions induced by LHW power in a given parallel velocity space, since the main mechanism of LHCD is Landau damping, which can be written as
\begin{eqnarray} \label{LHW operator}
      \hat{D}_{\mathrm{LH}} \equiv   \frac{u_{e}^2}{m_{e} T_{e}} \frac{\partial}{\partial u_{\parallel}} D_{\mathrm{LH}}\left(u_{\parallel}\right) \frac{\partial}{\partial u_{\parallel}} \, , 
\end{eqnarray}
with
\begin{eqnarray} 
     D_{LH}\left(u_{\parallel}\right) = 
    \begin{array}{ll}
        0 \, ,  & u_{\parallel} < u_{\parallel, 1} \, ,\\
	D_{LH}  \, , & u_{\parallel, 1} \leq u_{\parallel} \leq u_{\parallel, 2} \, ,\\
	0 \, , & u_{\parallel}>u_{\parallel, 2} \, ,
    \end{array}
\end{eqnarray}
where $D_{LH}$ is a constant determined by the wave power~\cite{Dumont2000combined}. This kind of approximation form of LHW diffusion operator implies that the influence of relativistic effects on the LHW resonance condition has been ignored; this is reasonable since the relativistic correction of LHW has a minor effect on SCD. Due to the accessibility of LHW, the parallel refractive index of LHW is upper to $2$, meaning the $u_{\parallel,1}$ should be larger than $2u_{e}$ at least according to the resonance condition of LHW.
% which will cause the and we use $D_{LH}$ in following derivations for convenience.}
In addition, $\hat{D}_{LH}$ is a self-adjoint operator (i.e., $\hat{D}_{LH}^\dagger = \hat{D}_{LH}$).

Combined the two operators $\hat{C^{\dagger}}$ and $\hat{D}_{LH}^\dagger$, the Eq.(\ref{general equation satisfy the response function}) can be rewritten as 
\begin{equation} \label{eq: equation of the response function in collision and diffusion caused by LHW}
	-v_{\parallel} \hat{b} \cdot \nabla \chi- \hat{C}^{\dagger}\chi - \hat{D}_{LH}^{\dagger} \chi=\frac{v_{\parallel} B}{\left\langle B^{2}\right\rangle}.
\end{equation}
The problem of solving SCD efficiency is reduced to obtain the analytical form of dimensionless response function $\tilde{\chi}$ corresponding to $\chi$ through Eq.(\ref{the relation between dimensionless response function and response function}), and then compute the integral Eq.(\ref{eq: general CD efficiency}) with given $S_{\mathrm{ECW}}(f_{M})$. The following subsections will be arranged to acquire the dimensionless response function $\tilde{\chi}$ by using perturbation and Green-Function techniques, and then to calculate the SCD efficiency versus the ECCD only case.

\subsection{The response function}
The assumption in ECCD~\cite{LinLiu2003} that the effective collisional frequency is much smaller than the bounced frequency, i.e., $(\nu_{e 0}/\epsilon)/\omega_{b} \ll 1$ where $\epsilon$ is the inverse aspect ratio, is also adopted, such that the trapped electrons are allowed to complete the banana orbits at all energies. Therefore, the response function $\chi$ has an expansion in this parameter (i.e., $\chi = \chi_{I} + \chi_{II} + ...$), for passing electrons Eq.(\ref{eq: equation of the response function in collision and diffusion caused by LHW}) can be writen as
\begin{equation} \label{eq: eq that satisfied passing electrons in view of response function}
	\frac{1}{v}\left\langle\frac{B}{|\xi|}( \hat{C}^{\dagger} + \hat{D}_{LH}^{\dagger})\right\rangle \chi_{I}=-\mathrm{sgn}(v_{\parallel}) .
\end{equation}
The subscript $I$ of $\chi_{I} $ will be omitted in the following derivations for articulating due to the limitation of the banana regime.   

According to the previous assumption that collisions still dominate the process of electron relaxation, although diffusions caused by LHW affect electrons' dynamic behavior in a given parallel velocity space,  we can linearize $\chi$  
\begin{equation} \label{eq: linearized response function}
	\chi = {\chi}_{0} + {\delta\chi} \, , 
\end{equation}
in which $D_{LH,0}\equiv \frac{D_{LH}}{m_{e} T_{e} \nu_{e0}}$ is the normalized, small quantity, representing the LH wave intensity compared to collisions~\cite{Dumont2004theory}. Substituting Eq.(\ref{eq: linearized response function}) into Eq.(\ref{eq: eq that satisfied passing electrons in view of response function}), by using the perturbation technique, we obtain the zero-order equation
\begin{equation} \label{eq: zero-order equation}
	\frac{1}{v}\left\langle\frac{B}{|\xi|} \hat{C}^{\dagger}\right\rangle \chi_{0} = - \mathrm{sgn} (v_{\parallel} ) \, ,
\end{equation}
and the first-order equation \label{eq: first-order equation}
\begin{equation}
	\langle  \frac{B}{|\xi|} \hat{C}^{\dagger} \rangle {\delta\chi} = - \langle \frac{B}{|\xi|} \hat{D}_{LH}^{\dagger} \rangle {\chi}_{ 0} \, .
\end{equation}

Lin-Liu~\cite{LinLiu2003} has given a useful solution for dimensionless response function $\tilde{\chi}_{0}$ corresponding to $\chi_{0}$ by separating variables $u,\lambda$ and using Green-Function technique
\begin{equation} \label{zero-order dimensionless response function}
	\tilde{\chi}_{0} = \mathrm{sgn} (u_{\parallel} ) F(u) H(\lambda) ,
\end{equation}
where $	\lambda = \frac{\mu B_{m}}{mu^{2}/2} = \frac{B_{m}}{B} \frac{u_{\perp}^{2}}{u^2} $ is a new pitch-angle variable. For passing electrons, $\lambda$ is in range of $0 \le \lambda < 1$, and for trapped particles $1 \le \lambda < \frac{B_{m}}{B}$, at a given poloidal angle. The more details of Eq.(\ref{zero-order dimensionless response function}) are in \ref{apd: Details of Lin-Liu's response function}.

Now, the specific form of the first-order equation associated with the dimensionless response function $\Tilde{\delta\chi}$ is
\begin{equation} \label{the specific form of first-order equation}
\fl	(\nu_{e i} + \nu_{D}) \bar{L} \tilde{\delta\chi} - \lambda_{s}(u) \langle \frac{B}{B_{m}} \frac{1}{|\xi|}   \frac{\partial}{\partial u} \tilde{\delta\chi} \rangle
	 = -  u_{e}^{2} \langle \frac{B}{B_{m}} \frac{1}{|\xi|}  \frac{\partial}{\partial u_{\parallel}} D_{LH,0}\nu_{e0} \frac{\partial}{\partial u_{\parallel}}  \tilde{\chi}_{0} \rangle ,
\end{equation}
where $\bar{L}$ is the bounce averaged pitch-angle scattering operator
\begin{equation}
	\bar{L} = \langle \frac{B}{B_{m}} \frac{1}{|\xi|} L \rangle = 2 \frac{\partial}{\partial \lambda} \lambda \langle |\xi| \rangle \frac{\partial}{\partial \lambda}.
\end{equation}
In order to obtain the analytical form of $\tilde{ \delta \chi }$, firstly we approximate the zero-order response function $\tilde{\chi}_{0}$ in the right side of Eq.(\ref{the specific form of first-order equation}) by keeping the first term of its Legendre expansion in $\xi$ 
\begin{eqnarray}
	\langle \frac{B}{B_{m}} \tilde{\chi}_{0} \rangle & \simeq \langle \frac{B}{B_{m}}  p_{1}(\xi) \frac{3}{2} \int_{-1}^{1} d\xi \tilde{\chi}_{0}(u, \xi, \theta_{p}) p_{1}(\xi) \rangle \cr
	& = \langle p_{1}(\xi) \rangle \frac{3}{2} \langle \frac{B^{2}}{B_{m}^{2}}  \rangle \int_{0}^{\frac{B_{m}}{B}} d\lambda sgn(u_{\parallel}) \tilde{\chi}_{0}(u,\lambda)  = f_{c} F(u) \langle \xi \rangle ,
\end{eqnarray}
with the fact $p_{1}(\xi) = \xi$ and $f_c$ is the effective circulating particle fraction in the neoclassical transport theory~\cite{fc_1995, LinLiu2003}. By substituting it into the right-hand side of Eq.(\ref{the specific form of first-order equation}) and using
\begin{equation}
	\frac{\partial}{\partial u_{\parallel}} = \xi \frac{\partial}{\partial u} + \frac{1}{u}(1-\xi^2)\frac{\partial}{\partial \xi} \, , 
\end{equation}
we have 
\begin{eqnarray} \label{deal with LHW operator acting on the zero-order response function}
    \fl	  u_{e}^{2} \langle  \frac{1}{ |\xi| } \frac{ \partial }{ \partial u_{\parallel}} D_{LH,0}\nu_{e0} \frac{\partial}{ \partial u_{\parallel}} ( f_{c} F(u) \xi ) \rangle \cr
\fl    = \mathrm{sgn}(u_{\parallel}) D_{LH,0}\nu_{e0} u_{e}^{2} f_{c} \bigg\langle \left[ \xi^2 \frac{\partial^2}{\partial u^2} F(u) + \frac{3 (1 - \xi^2)}{u} \frac{\partial F(u)}{\partial u} +  \frac{-3  (1 - \xi^2) }{u^2} F(u)  \right] \bigg\rangle \, .
\end{eqnarray}
For the convenience of calculation, we numerically approximate $F(u)$ as a simple function with fixed order
\begin{equation} \label{approximate F(u) by a simple function}
	F(u) = C_{k} ( \frac{u}{u_{e}} )^{k},
\end{equation}
where $C_{k}$ is a coefficient and $k$ is the fixed order. The specific form of $F(u)$ and its derivatives we adopted in the following derivations and discussions on their effectiveness are in \ref{apd: The approximation of Fu}. Substituting Eq.(\ref{approximate F(u) by a simple function}) into Eq.(\ref{deal with LHW operator acting on the zero-order response function}), then Eq.(\ref{the specific form of first-order equation}) reduces to
\begin{equation} \label{the explicit form of first-order eq after approximate F(u)}
	\fl 	( \nu_{ei} + \nu_{D} ) \bar{L} \tilde{ \delta \chi } - \lambda_{s}(u) \langle \frac{B}{B_{m}}\frac{1}{ |\xi| } \rangle  \frac{ \partial }{ \partial u} \tilde{ \delta \chi }  =  - \mathrm{sgn}(u_{\parallel}) D_{LH,0}\nu_{e0} A_{1} \frac{u^{k-2}}{u_{e}^{k-2}} D_{k}(\lambda) ,
\end{equation}
where
\begin{equation}
	D_{k}(\lambda) = \langle \left[ (k-3)\xi^2 + 3 \right] \rangle = (3-k)\lambda h_{1} + k,
\end{equation}
with $ A_{1} = (k-1) C_{k} f_{c} $ and $h_{1} = \langle \frac{B}{B_{m}} \rangle$.
%we obtain
%\begin{eqnarray}
% \fl 	\nu_{e 0} u_{e}^{2} \langle \frac{1}{ |\xi| } \frac{ \partial }{ \partial u_{\parallel}} D_{LH} \frac{\partial}{ \partial u_{\parallel}} ( f_{c} F(u) \xi) \rangle  =  sgn(u_{\parallel}) \nu_{e 0}  D_{LH} A_{1} \frac{u^{k-2}}{u_{e}^{k-2}} \langle \left[ (k-3)\xi^2 + 3 \right] \rangle,
%\end{eqnarray}
%with 
We suppose the perturbation response function still can be separated by variables $u$ and $\lambda$, which results in a useful form
\begin{equation}
	\frac{\tilde{ \delta \chi }}{D_{k}(\lambda)} = \tilde{ \delta \chi }^{\prime} = \mathrm{sgn}(u_{\parallel}) S(u) M(\lambda).
\end{equation}
Likely, we adopt the same technique to approximate the slowing-down part of Eq.(\ref{the explicit form of first-order eq after approximate F(u)})
\begin{eqnarray}
\lambda_{s}(u)	\langle   \frac{B}{B_{m}}\frac{1}{ |\xi| }   \frac{ \partial }{ \partial u} \tilde{ \delta \chi }^{\prime} \rangle & \simeq \mathrm{sgn}(u_{\parallel}) \lambda_{s}(u)  \frac{3}{2} \langle \frac{B^{2}}{B_{m}^2} \rangle  \int_{0}^{1} d\lambda   M(\lambda) \frac{ \partial  S(u) }{ \partial u} \cr
 	& = \mathrm{sgn}(u_{\parallel})f_{n} \lambda_{s}(u) \frac{ \partial  S(u) }{ \partial u},
\end{eqnarray}
where
\begin{equation}
	f_{n} = \frac{3}{2} \langle \frac{B^{2}}{B_{m}^2} \rangle  \int_{0}^{1} d\lambda   M(\lambda).
\end{equation}
% in which trapped particles are not included in the integral.
%Thus the first-order equation becomes
% \begin{equation}
% \fl   \frac{1}{D_{k}(\lambda)} ( \nu_{ei} + \nu_{D} ) S(u) \bar{L} D_{k}(\lambda) M(\lambda)  -  \lambda_{s}(u) f_{n} \frac{\partial}{\partial u} S(u) = - D_{LH} A_{1} \frac{u^{k-2}}{u_{e}^{k-2}} .
% \end{equation}
Furthermore, the pitch-angle scattering term about $\lambda$ satisfies
\begin{equation}
	\bar{L} D_{k}(\lambda)M(\lambda) = - D_{k}(\lambda) ,
\end{equation} 
With the regular boundary condition at $\lambda = 0$ and a new definition $E_{k}(\lambda) \equiv D_{k}(\lambda)M(\lambda)$, we obtain
\begin{equation}
	E_{k}(\lambda) = \frac{\theta(1 - \lambda)}{2} \int_{\lambda}^{1} d\lambda^{\prime} \frac{D_{k}(\lambda^{\prime} )}{ \langle ( 1-\lambda^{\prime}B/B_{m} )^{1/2} \rangle } ,
\end{equation}
the more useful forms about $E_{k}(\lambda)$ are in \ref{apd: The interpolation formulation of E_lambda}.

For now, we can separate the equation about $u$ 
\begin{equation}
	  ( \nu_{ei} + \nu_{D} ) S(u) +  \lambda_{s}(u) f_{n} \frac{\partial}{\partial u} S(u) =   D_{LH,0}\nu_{e0} A_{1} \frac{u^{k-2}}{u_{e}^{k-2}}.
\end{equation}
Substituting coefficients and using the Green-Function technique, where the Green function is
\begin{equation}
	G(u, u^{\prime}) = \theta(u - u^{\prime})  (\frac{u^{\prime}}{u})^{\rho^{\prime}} (\frac{\gamma + 1}{\gamma^{\prime} + 1})^{\rho^{\prime}},
\end{equation}
with the boudary condition $S(0) = 0$, we find the relativistic form of $S(u)$
% \begin{eqnarray}
% 	S_{r}(u) & = D_{LH} A_{2} \frac{1}{u_{e}} ( \frac{\gamma + 1}{u} )^{\rho^{\prime}} \int_{0}^{u} d u^{\prime} ( \frac{u^{\prime}}{\gamma^{\prime} + 1} )^{\rho^{\prime}} ( \frac{u^{\prime}}{u_{e}} )^{k} ( \frac{1}{\gamma^{\prime}} )^2 \cr
% 	& = D_{LH} A_{2} (\gamma + 1)^{\rho^{\prime}} ( \frac{u}{u_{e}} )^{k+1} \int_{0}^{1} dy y^{\rho^{\prime} + k} \frac{1}{( \gamma^{\prime}_{y} + 1)^{\rho^{\prime}}} \frac{1}{( \gamma^{\prime}_{y} )^2}, \quad 
% \end{eqnarray}
% where $\rho^{\prime} = \frac{Z_{eff}+1}{f_{n}}$, $\gamma^{\prime}_{y} = \sqrt{ 1 + (uy/c)^2 }$, and $	A_{2} = \frac{A_{1}}{f_{n}} = (k-1)C_{k}\frac{f_{c}}{f_{n}}$.
% \cjn{
% \begin{eqnarray} \label{S(u)}
% 	S(u) & = D_{LH,0} A_{2} \frac{1}{u_{e}} ( \frac{\gamma + 1}{u} )^{\rho^{\prime}} \int_{0}^{u} d u^{\prime} ( \frac{u^{\prime}}{\gamma^{\prime} + 1} )^{\rho^{\prime}} ( \frac{u^{\prime}}{u_{e}} )^{k} ( \frac{1}{\gamma^{\prime}} )^2 \cr
% 	& = D_{LH,0} A_{2} (\gamma + 1)^{\rho^{\prime}} ( \frac{u}{u_{e}} )^{k+1} \int_{0}^{1} dy y^{\rho^{\prime} + k} \frac{1}{( \gamma^{\prime}_{y} + 1)^{\rho^{\prime}}} \frac{1}{( \gamma^{\prime}_{y} )^2}, \quad 
% \end{eqnarray}
% }
\begin{eqnarray} \label{S(u)}
	S(u) & = D_{LH,0} A_{2} \frac{1}{u_{e}} ( \frac{\gamma + 1}{u} )^{\rho^{\prime}} \int_{0}^{u} d u^{\prime} ( \frac{u^{\prime}}{\gamma^{\prime} + 1} )^{\rho^{\prime}} ( \frac{u^{\prime}}{u_{e}} )^{k} ( \frac{1}{\gamma^{\prime}} )^2 \cr
	& = D_{LH,0} A_{2} ( \frac{u}{u_{e}} )^{k+1} \int_{0}^{1} dy y^{\rho^{\prime} + k} \left( \frac{\gamma + 1}{ \gamma^{\prime}_{y} + 1} \right)^{\rho^{\prime}} \frac{1}{( \gamma^{\prime}_{y} )^2}, \quad 
\end{eqnarray}
where $\rho^{\prime} = \frac{Z_{eff}+1}{f_{n}}$, $\gamma^{\prime}_{y} = \sqrt{ 1 + (uy/c)^2 }$, and $	A_{2} = \frac{A_{1}}{f_{n}} = (k-1)C_{k}\frac{f_{c}}{f_{n}}$.

To sum up, we have the semi-analytical relativistic response function $\tilde{ \delta \chi }$ to describe the perturbation of electron relaxation caused by LHW effect in a given parallel velocity space
\begin{equation}
	\tilde{ \delta \chi } = \mathrm{sgn}(u_{\parallel}) S(u) E(\lambda).
\end{equation}
Although diffusions induced by LHW power can influence electron's behavior outside of a given region through pitch-angle scattering revealed by kinetic simulation, this effect will exponentially decay so that we can esteem $\tilde{ \delta \chi }$ only appear in the given region, which is consistent with the fact that $D_{LH}$ equals zero outside of a given region.

\subsection{The SCD efficiency}
According to previous discussions, the term $S_{rf}(f_{M})$ characterizing properties of ECW is necessary to calculate the SCD efficiency. Following Refs.~\cite{Cohen1987, LinLiu2003}, the simplified quasilinear diffusion operator for ECW is
\begin{equation}
	S_{ECW}(f) = \delta(x - x_{R}) \tilde{\Lambda} D_{ECW} \delta( \omega - k_{\parallel}v_{\parallel} - \ell \frac{\omega_{c}}{\gamma} ) \tilde{\Lambda} f ,
\end{equation}
with $\tilde{\Lambda} = \frac{\partial}{\partial w} + \frac{k_{\parallel}}{\omega} \frac{\partial}{\partial p_{\parallel}}$. Here, $x_{R}$ denotes the spatial location of ECW power deposition, the other symbols $\omega, \omega_{c}, k_{\parallel}, \ell$, respectively, are the frequency of ECW, the local cyclotron frequency, the parallel wave number of ECW, and the harmonic number.

The quasilinear diffusion coefficient $D_{ECW}$ is widely used in linear and quasilinear theory, as well as numerical simulations, whose general form is given by~\cite{Cohen1987}
    \begin{eqnarray} \label{D}
        D_{ECW} = \pi e^{2} |U_{\ell}|^{2} \, ,
    \end{eqnarray}    
    with
    \begin{eqnarray} \label{electric field part}
        U_{\ell} = E_{\parallel}v_{\parallel} J_{\ell-1}(a) + 2^{-1/2} v_{\perp} \left[ E_{-}e^{-i\psi}J_{\ell}(a) + E_{+}e^{i\psi}J_{\ell+1}(a) \right] \, ,
    \end{eqnarray}
    where $E_{+}$, $E_{+}$~(defined as $E_{\pm} = E_{x} \pm i E_{y}$) and $E_{\parallel}$, respectively, are the left-hand, right-hand circularly polarized, and parallel (to $\mathbf{B}$) components of the electric field. The $J_{\ell}$ with different subscripts are the Bessel functions, and $\psi$ is the angle between the wave vector and the axis. 
    % The velocity components are evaluated at the resonance position. The variable $a$ is defined as $a = k_{\perp} \rho$, where $k_{\perp}$ is the perpendicular wavenumber and $\rho$ is the electron gyroradius. In general, $a$ is assumed to be sufficiently small so that the Bessel functions can be expanded, then Eq.(\ref{electric field part}) can be approximated further.
    Notably, the original form of Eq.(\ref{D}) in \cite{Cohen1987} has a relativistic factor $\gamma$~[see Eq.(13a) therein]. The reason that we omit $\gamma$ here is due to the different definitions of the absorbed rf power density $Q$ between Lin-Liu's model~\cite{LinLiu2003} [see Eq.(9) and Eq.(35) therein] and Cohen's model~\cite{Cohen1987} [see Eq.(10) therein]. Here, we adopt the former definition for consistency. Both models use the second harmonic mode~(X2-mode) of ECW to calculate ECCD efficiency, whose ECW coefficient $D_{ECW}$ is approximated by
    \begin{eqnarray}
        D_{X2} = \frac{ \pi e^{2} |E_{-}|^{2} }{2} \left( \frac{k_{\perp} u_{\perp}}{ 2 \omega_{c} } \right)^{2\ell-2} \frac{u_{\perp}^{2}}{\gamma^{2}} \, ,
    \end{eqnarray}
    with $\ell = 2$.
For benchmarking with ECCD's result~\cite{LinLiu2003}, the X2-mode of ECW is adopted in the following derivations of this section. The O1-mode of ECW, widely used in current reactors, is employed to show SCD results with higher parameters in Sec.\ref{sec: the SCD results with parameters used in current reactors}.
% to show results of CD efficiency in the following de}

According to the previous assumption about power deposition, following~\cite{LinLiu2003}, the SCD efficiency can be written as
\begin{equation}\label{eq: the explicit form of SCD efficiency}
	\zeta^{*}_{SCD}=-\frac{4}{\ln \Lambda}\left\langle\frac{B}{B_{m }}\right\rangle \frac{m u_{e}^{2} \int d \gamma\left(u_{\perp}\right)^{2 \ell} f_{M} \tilde{\Lambda} \tilde{\chi}}{\int d \gamma\left(u_{\perp}\right)^{2 \ell} f_{M}} ,
\end{equation}
where $u_{\perp}^{2} = c^{2} \left[ \gamma^2 - 1 - ( \frac{\gamma - y}{n_{\parallel}} )^2 \right] $, $n_{\parallel} = k_{\parallel}c/\omega$ and $y = \ell \omega_{c}/\omega$. The upper $\gamma_{max}$ and lower $\gamma_{min}$ bounds of integrals are determined by 
\begin{equation} \label{eq: the bounds of ECW resonance}
	\gamma^2 - 1 - ( \frac{\gamma - y}{n_{\parallel}} )^2 = 0.
\end{equation}
Using the linearized response function, we have
\begin{equation} \label{eq: m ue linearized with response function}
	m u_{e}^{2} \tilde{\Lambda} \tilde{\chi} = m u_{e}^{2} \tilde{\Lambda}( \tilde{\chi}_{0} + \tilde{ \delta \chi } ),
\end{equation}
with
\begin{eqnarray}\label{eq: m ue with perturbation response function}
	\fl   m u_{e}^{2} \widetilde{\Lambda} \widetilde{\delta\chi} & = & \gamma \frac{u_{e}^{2}}{u} \frac{\partial}{\partial u} \widetilde{\delta\chi}+2 \frac{B_{\max }}{B} \frac{u_{\|} u_{e}^{2}}{u^{3}}\left(\frac{\gamma u_{\|}}{u}-\frac{n_{\|} u}{c}\right) \frac{\partial}{\partial \lambda} \widetilde{\delta\chi} \cr 
	 & = & sgn \left(u_{\|}\right)\left\{\gamma \frac{u_{e}^{2}}{u} \frac{d S}{d u} E + 2 \frac{B_{\max }}{B} \frac{u_{\|} u_{e}^{2}}{u^{3}}\right. \left.\times\left(\frac{\gamma u_{\|}}{u}-\frac{n_{\|} u}{c}\right) S \frac{d E}{d \lambda}\right\} \, ,
\end{eqnarray}
and the term $m u_{e}^{2} \tilde{\Lambda}\tilde{\chi}_{0}$ representing the contribution of ECCD has been given in~\cite{LinLiu2003}.
Thus the Eq.(\ref{eq: m ue linearized with response function}) actually implies
\begin{equation} \label{eq: linearized CD efficiency}
	\zeta^{*}_{SCD} = \zeta^{*}_{ECCD} + \zeta^{*}_{syn}.
\end{equation}
% The  in Eq.(\ref{eq: m ue linearized with response function}) and $\zeta_{ECCD}^{*}$ in Eq.(\ref{eq: linearized CD efficiency}) both have been given by Lin-Liu~\cite{LinLiu2003}. 

The $\zeta^{*}_{SCD}$ is a function of electron temperature $T_{e}$, parallel refractive index $n_{\parallel}$, effective charge $Z_{eff}$, the parameter $y = \ell \omega_{c}/\omega$, the inverse aspect ratio $\varepsilon$, the poloidal angle $\theta_{p}$ in a given flux, the constant $D_{LH}$ determined by the LHW power, and the given parallel velocity boundaries $u_{\parallel,1}$ and $u_{\parallel,2}$.

\section{Results and discussion} \label{sec: results}
\subsection{Theoretical results of SCD efficiency} 
\begin{figure}
    \centering
    \includegraphics[width=0.5\linewidth]{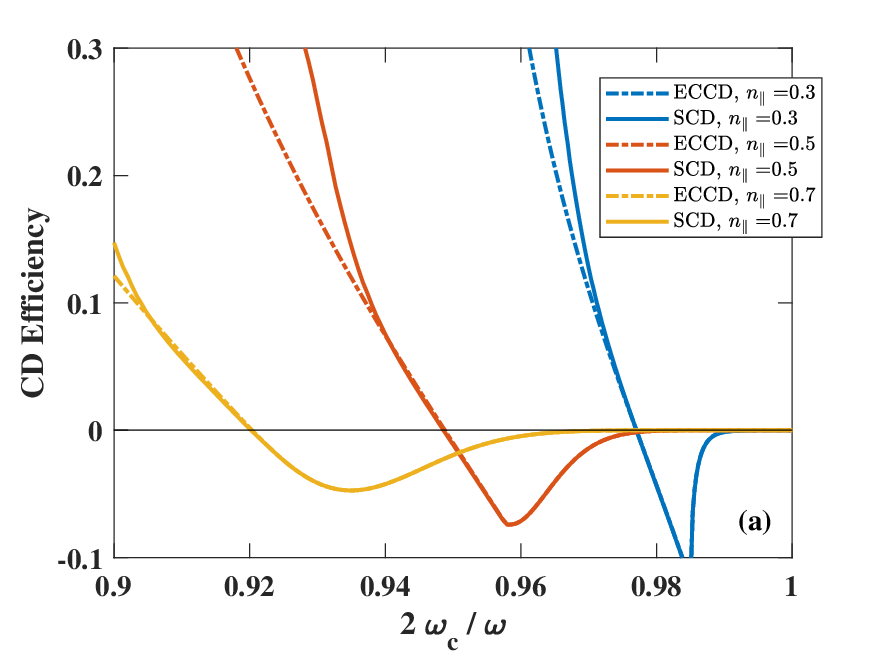}
    \includegraphics[width=0.5\linewidth]{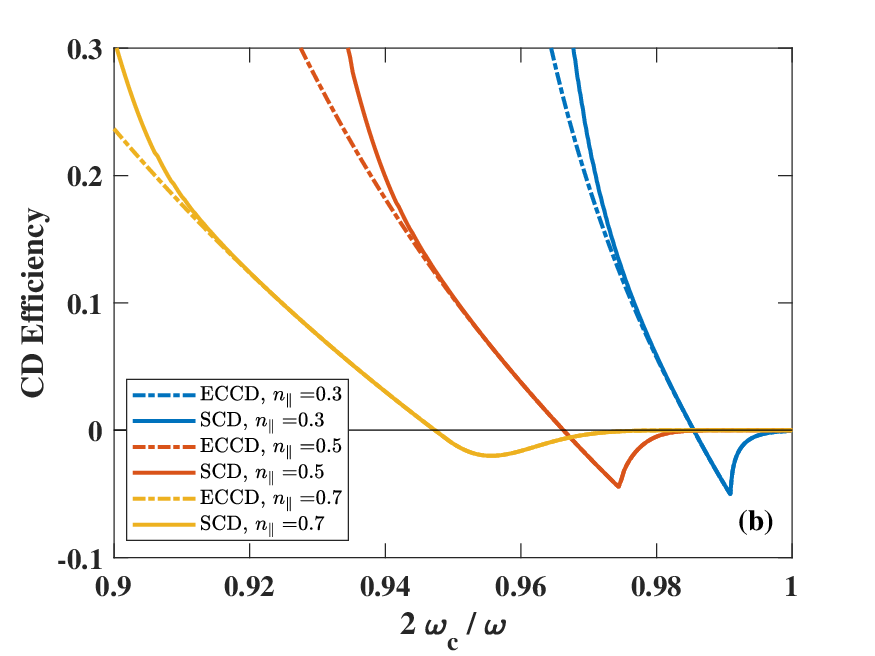}
    \includegraphics[width=0.5\linewidth]{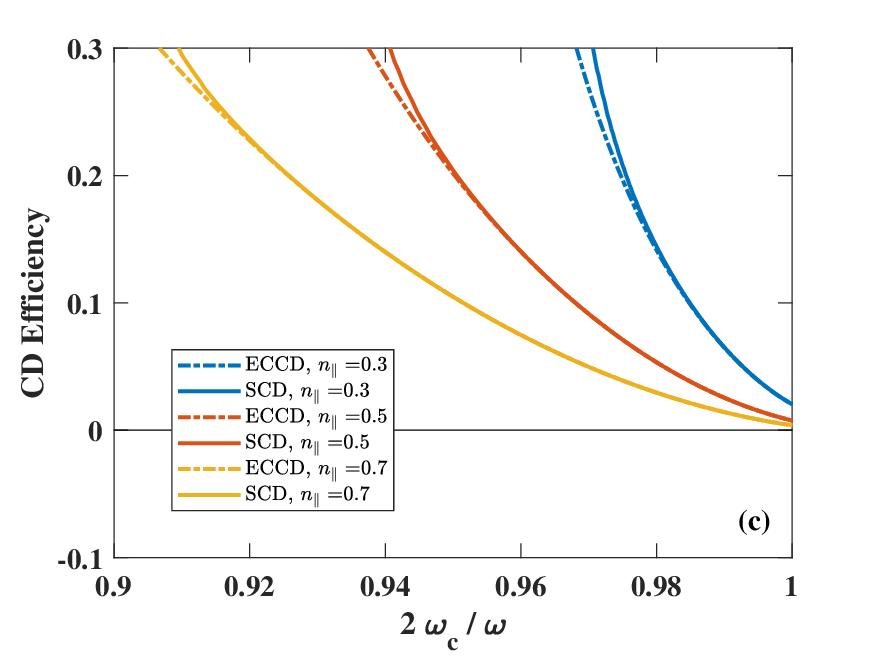}
    \caption{The ECCD(dashed lines) and SCD(solid lines) efficiency at different poloidal angles (a)$\theta_{p} = 15^{\circ}$(weak-field side), (b) $\theta_{p} = 90^{\circ}$ and (c)$\theta_{p} = 165^{\circ}$ at a given flux surface. The color of the line denotes the different parallel refractive index of ECW.}
    \label{fig:SCD result benchmark with ECCD}
\end{figure}

For now the $\zeta^{*}_{SCD}(Te, n_{\parallel}, Z_{eff}, y, \varepsilon, \theta_{p}, D_{LH}, u_{\parallel,1}, u_{\parallel,2})$  can be calculated according to the above derivations.
% though Eq.(\ref{},\ref{eq: the explicit form of SCD efficiency}-\ref{eq: m ue with perturbation response function}). 
To benchmark SCD's results with ECCD’s~\cite{LinLiu2003}, we take the same plasma and magnetic field parameters as the ECCD case; these parameters and LHW parameters are shown in Tab.~\ref{tab: Plasma, magnetic field and LHW parameters in ECCD case}, and the resonant mechanism of ECW is X2-mode. The dimensionless SCD efficiency $\tilde{\zeta}_{SCD}$ and ECCD efficiency $\tilde{\zeta}$ as a function of $y = \frac{2 \omega_{c}}{\omega}$ with different poloidal angle $\theta_{p}$ are depicted as shown in Fig.\ref{fig:SCD result benchmark with ECCD}.
\begin{table}[htbp]
	\centering
	\caption{Plasma, magnetic field, and LHW parameters}
	\begin{tabular}{c c}
		\hline
		$T_{e}$ & $2000 \, \mathrm{eV}$ \\
            $n_{e}$ & $2*10^{19} \, \mathrm{m}^{-3}$ \\
		$n_{\parallel} $ & $0.3, 0.5, 0.7$ \\
		$Z_{eff}$ & 1.6 \\
		$\epsilon$ & $0.2$ \\
		$\theta_{p}$ & $ \frac{\pi}{12}, \frac{\pi}{2}, \frac{11\pi}{12}$ \\
		% toroidal field & $ B = B_{0}/(1+\varepsilon cos\theta_{p}) $ \\
		% poloidal field & $ B = B_{p0}/(1+\varepsilon cos\theta_{p}) $ \\
		R & $1.76 \, \mathrm{m}$ \\
		\hline
		$D_{LH,0}$ & $0.2$ \\
		$u_{\parallel,1}$ & $2u_{e}$ \\
		$u_{\parallel,2}$ & $4u_{e}$ \\
		\hline
	\end{tabular}
        \label{tab: Plasma, magnetic field and LHW parameters in ECCD case}
\end{table}

% \begin{figure}[htbp] \label{SCD results benchmark with ECCD}
% 	\centering
% 	\subfigure{
% 	\includegraphics[width=0.55\linewidth]{picture/EffTp15.eps}
% 	}
% 	\subfigure{
% 	\includegraphics[width=0.55\linewidth]{picture/EffTp90.eps}
% 	}
% 	\subfigure{
% 	\includegraphics[width=0.55\linewidth]{picture/EffTp165.eps}
% 	}
% 	\caption{The ECCD(dashed lines) and SCD(solid lines) efficiency at different poloidal angles (a)$\theta_{p} = 15^{\circ}$(weak-field side), (b) $\theta_{p} = 90^{\circ}$ and (c)$\theta_{p} = 165^{\circ}$ at a given flux surface. The color of the line denotes the different parallel refractive index of ECW.}
% \end{figure}
The three panels indicate the existence of the improvement of ECCD efficiency in the presence of LHW (i.e., the synergy effect) in certain parameter spaces. Comparative analysis between the results of ECCD and SCD reveals two distinct features. One is that the threshold value of $y$, at which the synergy effect becomes sufficiently significant, shifts towards higher values as $n_{\parallel}$ decreases. The feature is due to the limitation relationship between $y$ and $n_{\parallel}$, $y \ge \sqrt{1 - n_{\parallel}^2}$, such that a smaller $n_{\parallel}$ will have a larger $y_{min}$. The other important feature is that the synergy effect is inclined to occur in smaller y with fixed $n_{\parallel}$. This feature is consistent with the previous understanding of SCD's mechanism based on the kinetic simulation~\cite{PhysRevLett.93.255002, Dumont2004theory}: the overlap of quasilinear domains of ECW and LHW in velocity space, which is a matching pattern between the resonance curve of ECW and the diffusion region of LHW.

\begin{figure}
    \centering
    \includegraphics[width=0.7\linewidth]{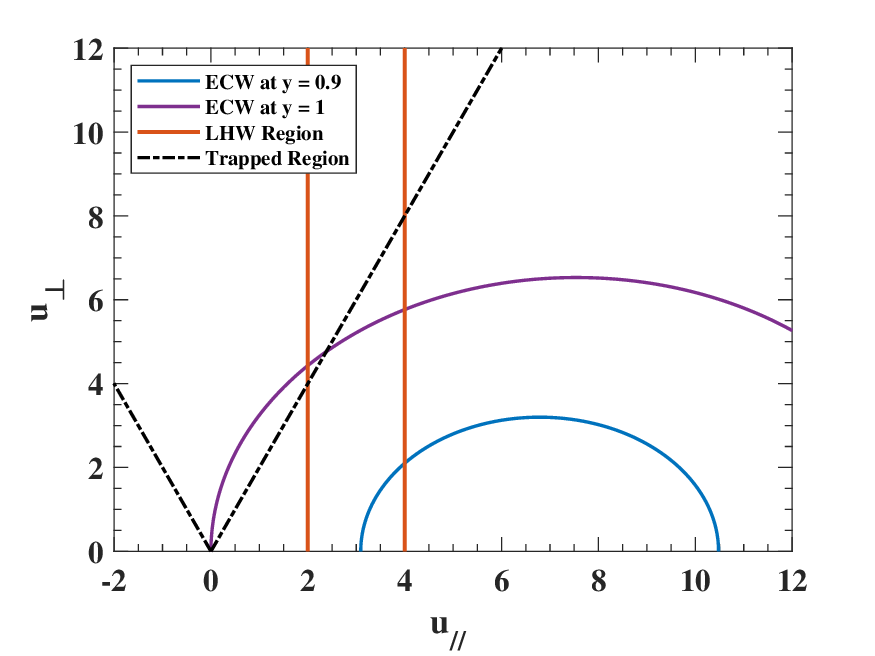}
    \caption{Resonance curves of ECW(the blue line: $y = 0.9$, the purple line: $y = 1$); $n_{\parallel}$ is fixed at $0.5$. The area between the two orange lines represents the diffusion region of LHW, and the area within the two black dashed lines describes the trapped region, both in velocity space.}
    \label{fig: resonant curve n_para=0.5 theta_p=90}
\end{figure}

Thus, we select a typical case of the above situations with $\theta_{p} = 90^{\circ}$ and $n_{\parallel} = 0.5$ to explain two features, as shown in Fig.\ref{fig: resonant curve n_para=0.5 theta_p=90}.
Varying from $y = 0.9$ to $y = 1$, the resonance curve of ECW changes from the blue line to the purple line. The diffusion region of LHW is fixed with $u_{\parallel,1} = 2$ and $u_{\parallel,2} = 4$ (normalized to the thermal velocity). The area within the two black lines is the trapped region. For ECCD only, the efficiency changes from negative to positive as $y$ decreases because the low-energy part of the resonance curve of ECW (i.e., the smaller $\gamma$ part) migrates out of the trapped region, which means the transition of CD from the Ohkawa mechanism to the Fisch-Boozer mechanism.

For SCD with smaller $y$, the low-energy part of the resonance curve of ECW is more located in the fixed diffusion region of LHW. ECW will excite additional electrons located in the diffusion region of LHW, and these electrons will obtain more energy and smaller relaxation, so that the current is enhanced and the synergy effect is stronger. This physical process corresponds to the additional contribution of the perturbed response function $\tilde{\delta \chi}$. The trapped region plays a minor role in this process due to two reasons, one is that the low-energy part of the diffusion region of LHW (i.e., the smaller $u_{\parallel}$ and $u_{\perp}$ in a given velocity space) is far from the trapped region in general, the other is the banana regime assumption adopt here, which means the trapped effect is infinite so that all electrons will not run away from the trapped region, i.e., $\tilde{\chi} = 0$ for trapped particles that don't contribute to the integral of Eq.(\ref{eq: the explicit form of SCD efficiency}). 

Besides, the kinetic simulation~\cite{Dumont2004theory} points out that a few electrons can be pushed by ECW through the pitch-angle scattering and slowing-down process and then enter the diffusion region of LHW, but this linear model does not involve this indirect kinetic process. The perturbed response function $\tilde{ \delta \chi}$ only exists in a fixed velocity space due to the local LHW operator, which suggests that the synergy effect occurs directly at the overlap of the two waves' quasilinear domains in the linear model. 

% Besides, since $\tilde{ \delta \chi}$ is obtained from the Legendre expansion of the slowing-down part of the collision operator and LHW operator, and it exists only in a fixed velocity space, very few electrons are pushed by ECW and then enter the diffusion region of LHW through the slowing-down process, which implies that the synergy effect is more inclined to occur directly at the overlap of the two waves' quasilinear domains in the linear model.

Finally, the magnitude of $D_{LH}$ is power-dependent, with its specific value determined by the LHCD model. While different modes yield slightly different quasilinear values, this linear model focuses not on the absolute $D_{LH}$ but rather on the rescaled quantity $D_{LH,0} = D_{LH}/m_{e}T_{e}\nu_{e0}$, which is treated as a first-order expansion parameter of the response function. The constant $D_{LH,0}$ directly affects the magnitude of the synergy effect. We depict SCD efficiency at $\theta_{p} = 90^{\circ}$ and $n_{\parallel} = 0.5$ with different $D_{LH,0}$ in the range of $y = 0.93-0.95$, where the synergy effect is relatively obvious, as shown in Fig.\ref{fig: The SCD efficiency with different D_{LH}}. The result indicates that the SCD efficiency increases as the $D_{LH,0}$ increases compared with the fixed ECCD case. The reason lies in that a larger $D_{LH,0}$ directly represents a larger perturbation response function $\tilde{ \delta \chi }$ caused by the LHW power, and a larger $D_{LH,0}$ also implies a larger modification of distribution $f_{M}\phi$. 

%However, there needs to be a limitation on $f_{M}\phi$ to make the linear approximation applicable. Therefore, the premise of low RF power density can be further clarified. First, the modification distribution caused by the LHW power is smaller than the Maxwell distribution. The sum of both distributions will serve as the background distribution, and the ECW will act on this distribution; hence, second, the ECW power needs to be smaller than the LHW power.}

\begin{figure}[htbp]
	\centering
	\includegraphics[width=0.7\linewidth]{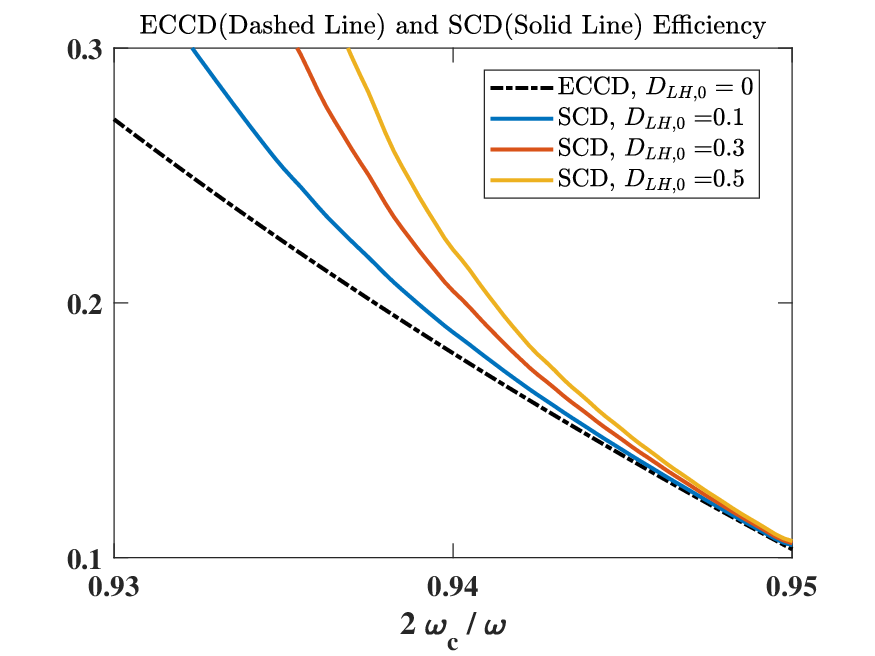}
	\caption{The SCD efficiency with different $D_{LH,0}$~(colored solid lines) based on fixed ECCD case~(black dashed line) at $\theta_{p} = 90^{\circ}$ and $n_{\parallel} = 0.5$.}
        \label{fig: The SCD efficiency with different D_{LH}}
\end{figure} 

\subsection{SCD without the trapping effect}

% Firstly, it's valuable to clarify the definition of a CD regime with or without the trapping effect. There are two aspects when considering this effect. One is to contemplate the situation in which the inverse aspect ratio is limited to zero~($\varepsilon \to 0$), which indicates that plasma tends to be a Lorentz gas in a uniform magnetic field and there are very few trapped electrons in calculations. This situation is defined as a regime without the trapping effect; the results and discussion in this subsection are mainly based on it. The other situation is that the poloidal angle at a given flux surface is limited to strong-field side~($\theta \to \pi$) where the trapped region tends to be extremely small, but the trapping effect is still taken into consideration. 

The pioneering work aiming to study the synergy effect theoretically~\cite{Dumont2004theory} hasn't taken the trapping effect~(i.e., the trapped electrons) into account since only the simplest Fisch-Boozer response function is applied in their model~[see Eq.(25) therein]. However, the trapping effect will influence the behaviour of resonance electrons; thus, investigating the impact of the trapping effect on SCD is significant to theoretical analysis.

The trapping effect influences SCD primarily through its modulation of ECCD in this linear model. Consequently, methods developed for studying ECCD without the trapping effect remain applicable to SCD investigations. The previous work~\cite{Start_1983} theoretically demonstrated that the ECCD efficiency without the trapping effect is larger than the case with the trapping effect~[see Fig.2-4 therein], it also indicated that the ECCD efficiency with the trapping effect is strongly dependent on the poloidal angle at a given flux surface. For numerical simulation, the optimized conditions of the synergy effect under the influence of the trapping effect were studied in~\cite{Yang2017, Wei_2016}, both of which utilized parameters from the EAST tokamak. For the ECCD model~\cite{LinLiu2003} that we are based on, by taking $\varepsilon \to 0$, the magnetic field $B/B_{max} \to 1$, and the effective passing particles fraction $f_{c} \to 1$, the response function $\Tilde{\chi}_{0}$ recovers the Fisch's results~\cite{Fisch1981_PhysRevA.24.3245}.

Investigating the impact of the trapping effect on SCD can start with the two aspects. In one aspect, the size of the trapping region at a given flux surface~($\varepsilon \neq 0$) is related to the magnitude of the poloidal angle, i.e., the position of the magnetic field where we calculate the CD efficiency. The three cases $\theta = \pi/12$~(weak-field side), $\theta = \pi/2$, and $\theta = 11\pi/12$ have been given in Fig.\ref{fig:SCD result benchmark with ECCD}. These results indicate that, for the same $n_{\parallel}$ and $y$, the CD efficiency increases as the poloidal angle $\theta_{p}$ increases, even transitioning from negative to positive values, which indicates a shift in the CD mechanism from the Ohkawa mechanism to the Fisch-Boozer mechanism. In another aspect, the same techniques similar to those in the ECCD model can also be applied in this linear model, resulting in the response function without the trapping effect
    \begin{eqnarray}
        \Tilde{\chi}_{wo} = \Tilde{\chi}_{0,wo} + \Tilde{\delta\chi}_{wo} \, ,
    \end{eqnarray}
    with
    \begin{eqnarray}
        \Tilde{\chi}_{0,wo} = \mathrm{sgn}(u_{\parallel})F_{wo}(u)H_{wo}(\lambda_{0}) \, ,
       \end{eqnarray}
       \begin{eqnarray}
         \Tilde{\delta\chi}_{wo} = \mathrm{sgn}(u_{\parallel}) S_{wo}(u) E_{k,wo}(\lambda_{0})  \, .
    \end{eqnarray}
    Here, $\lambda_{0} = u_{\perp}^{2}/u^{2}$, $H_{wo} = \sqrt{1 - \lambda_{0} }$, and $E_{k,wo} = \frac{\sqrt{1-\lambda_{0}}}{3} \left[ (3-k)\lambda_{0} + k + 6 \right]$, and the subscript $wo$ means ``without trapped electrons". The function $F_{wo}(u)$ had been given by~\cite{Fisch1981_PhysRevA.24.3245}, and $S_{wo}(u)$ can be written as 
\begin{eqnarray} \label{S(u)}
\fl	S_{wo}(u) = D_{LH,0} (k-1) C_{k} (\gamma + 1)^{\rho^{\prime}_{wo}} ( \frac{u}{u_{e}} )^{k+1} \int_{0}^{1} dy y^{\rho^{\prime}_{wo} + k} \frac{1}{( \gamma^{\prime}_{y} + 1)^{\rho^{\prime}}} \frac{1}{( \gamma^{\prime}_{y} )^2}, \quad 
\end{eqnarray}
with $\rho^{\prime}_{wo} = \frac{Z_{eff}+1}{f_{n,wo}}$,  and $f_{n,wo} \to (1/3 + 6/(k-3) - 6\sqrt{3} \arctan(\sqrt{k/3 - 1}) / (k-3)^{3/2} ) $. This form is useful for the following numerical calculations.
% in Eq. about velocity $F_{wo}(u)$ and $S_{wo}(u)$ can be obtained by taking $f_{c} \to 1$ and $f_{n} \to (1/3 + 6/(k-3) - 6\sqrt{3} \arctan(\sqrt{k/3 - 1}) / (k-3)^{3/2} )$ for $F(u)$~ and $S(u)$, and the subscript $wo$ means ``without trapped electrons".

% Furthermore, by taking the non-relativistic limitation $c \to \infty$, $k=4$ and $C_{k} = 1/(Z_{eff} + 5)$, $\Tilde{\chi}_{0,wo}$ is reduced to the nonrelativistic Fisch-Boozer response function
%     \begin{eqnarray}
%         \Tilde{\chi}_{0,wo,nr} = sgn(u_{\parallel}) \frac{1}{5 + Z_{eff}} (\frac{u}{u_{e}})^4 \xi \,, 
%     \end{eqnarray}
%     and $\Tilde{\delta\chi}_{wo,nr}$ is reduced to 
%     \begin{eqnarray}
% 		\Tilde{\delta\chi}_{wo,nr}  =  D_{LH,0} \frac{3}{( Z_{eff} + 5 )( Z_{eff} + 5.4595)} ( \frac{u}{u_{e}} )^5  \xi \left[ 3 + \frac{1}{3} \xi^{2} \right] \, .
%     \end{eqnarray}
  %    \begin{eqnarray}
		% \Tilde{\delta\chi}_{wo,nr}  =  D_{LH,0} \frac{3}{( Z_{eff} + 5 )( Z_{eff} + 1 + 5f_{n,wo} )} ( \frac{u}{u_{e}} )^5  \xi \left[ 3 + \frac{1}{3} \xi^{2} \right] \, .
  %   \end{eqnarray}
    % with  $\xi = u_{\parallel}/u$.

Correspondingly, the SCD efficiency without the trapping effect needs to be corrected, which is given by
    \begin{equation}
    	\zeta^{*}_{SCD,wo}=-\frac{4}{\ln \Lambda} \frac{m u_{e}^{2} \int \mathrm{d} \gamma\left(u_{\perp}\right)^{2 \ell} f_{M}       \tilde{\Lambda} \tilde{\chi}_{wo}}{\int \mathrm{d} \gamma\left(u_{\perp}\right)^{2 \ell} f_{M}} ,
    \end{equation}
    with 
    \begin{eqnarray}
    \fl m u_{e}^{2} \Tilde{\Lambda} \Tilde{\chi}_{0,wo} 
    	& = & \mathrm{sgn} \left(u_{\|}\right)\left\{ \gamma \frac{u_{e}^{2}}{u} \frac{\mathrm{d} F_{wo}}{\mathrm{d} u} H_{wo} + 2 \frac{u_{\|} u_{e}^{2}}{u^{3}} \left(\frac{\gamma u_{\|}}{u}-\frac{n_{\|} u}{c}\right) F_{wo} \frac{\mathrm{d} H_{wo}}{\mathrm{d} \lambda_{0}} \right\} \, , \\
    \fl	 m u_{e}^{2} \Tilde{\Lambda} \Tilde{\delta\chi}_{wo} 
    	& = & \mathrm{sgn} \left(u_{\|}\right)\left\{ \gamma \frac{u_{e}^{2}}{u} \frac{\mathrm{d} S_{wo}}{\mathrm{d} u} E_{k,wo} + 2 \frac{u_{\|} u_{e}^{2}}{u^{3}} \left(\frac{\gamma u_{\|}}{u}-\frac{n_{\|} u}{c}\right) S_{wo} \frac{\mathrm{d} E_{k,wo}}{\mathrm{d} \lambda_{0}} \right\} \, .
    \end{eqnarray}
    
\begin{figure}[htbp]
        \centering
        \includegraphics[width=0.5\linewidth]{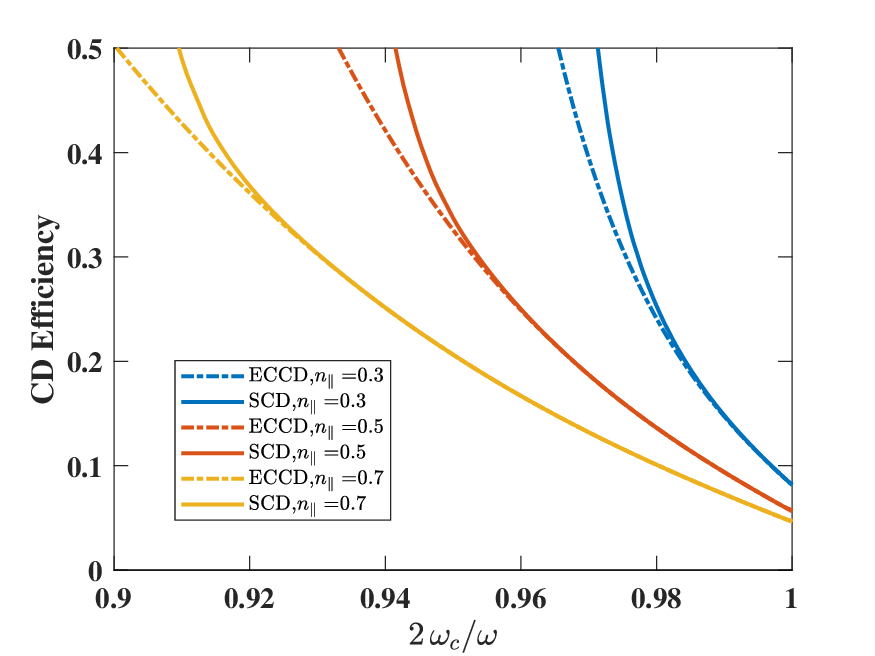}
        \caption{The ECCD~(dashed line) and SCD~(solid line) efficiency without the trapping effect~(by taking inverse aspect ratio $\varepsilon \to 0$). The parameters used here are the same as those in Tab.\ref{tab: Plasma, magnetic field and LHW parameters in ECCD case} for benchmarking with the ECCD model. }
        \label{fig:CDefficiency_without_trapped_electrons}
    \end{figure}

So far, the ECCD and SCD efficiency without the trapping effect can be calculated according to the above equations; the parameters used here are the same as those in Tab.\ref{tab: Plasma, magnetic field and LHW parameters in ECCD case} for benchmarking with ECCD model~\cite{LinLiu2003}. As shown in Fig.\ref{fig:CDefficiency_without_trapped_electrons}, the result indicates that there is no Ohkawa current, and the overall ECCD and SCD efficiency have increased compared to the cases with the trapping effect.

In addition, the relativistic effect will change the resonance curve of ECW and LHW in velocity space, and the synergy effect is virtually impossible to occur in the non-relativistic regime. But the response function in the non-relativistic limit can enhance our understanding of the theory. By taking the non-relativistic limit $c \to \infty$, the response function $\Tilde{\chi}_{0}$ recovers the Taguchi's result~\cite{Taguchi1983}. 

Furthermore, using the non-relativistic form of $F(u)$~\cite{LinLiu2003, Taguchi1983} with $k = 4$ and $ C_{k} = 1/(Z_{eff} + 1 + 4f_{c})$, the non-relativistic form of $S(u)$ can be written as 
\begin{eqnarray}\label{non-relavistic form of Su}
S_{nr}(u) = 3 D_{LH,0} \frac{f_{c}}{( Z_{eff} + 1 + 4f_{c} )( Z_{eff} + 1 + 5f_{n} )} ( \frac{u}{u_{e}} )^5 .
\end{eqnarray}
% \begin{eqnarray}\label{non-relavistic form of Su}
% 		S_{nr}(u) & = D_{LH,0} (k-1) \frac{ C_{k}}{ Z_{eff} + 1 + kf_{n}}  \frac{f_{c}}{f_{n}}  (\frac{u}{u_{e}})^{k+1} \cr
% 		& = 3 D_{LH,0} \frac{1}{( Z_{eff} + 1 + 4f_{c} )( Z_{eff} + 1 + 4f_{n} )} \frac{f_{c}}{f_{n}} ( \frac{u}{u_{e}} )^5 .
% \end{eqnarray}
The coefficients $f_{c}$ and $f_{n}$ indicate the rectification of the trapping effect on the perturbed response function. It's notable that the order of speed of perturbation response function, which is $(u/u_{e})^5$ revealed by Eq.(\ref{non-relavistic form of Su}), is consistent with that in~\cite{Dumont2004theory}~[see Eqs. (33-35) therein] by taking the Legendre polynomials $l=0$ and the same $D_{LH}(u_{\parallel})$.

\subsection{The results of the quasilinear simulation}

To examine the accuracy of our linear model, we use a two-dimensional Fokker-Planck code~\cite{shoucri1987application, SHOUCRI1989253} with the relativistic effect and the trapping effect to simulate ECCD and SCD. This model numerically solves the evolution of electron distribution with the effects of ECW and LHW simultaneously, which is given by
\begin{eqnarray}
    \frac{\partial f}{ \partial t} = (\frac{\partial f}{\partial t})_{ECW} + (\frac{\partial f}{\partial t})_{LHW} + (\frac{\partial f}{\partial t})_{Coll} \, ,
\end{eqnarray}
to compute the current density $j$ under the given injection power $P$. The CD efficiencies are defined by
\begin{equation} \label{eq: thee definition of ECCD in quasilinear simulation}
	\xi_{ECCD} = \frac{j_{ECW}}{P_{ECW}}, 
\end{equation}
\begin{equation} \label{eq: the definition of SCD in quasilinear simulation}
	\xi_{SCD} = \frac{j_{LHW+ECW} - j_{LHW}}{P_{ECW}}.
\end{equation}

In this section, the results of the linear model and the quasilinear simulation are compared to check the linear analysis and clarify its limitations. Notably, the key insight lies in the trend rather than the exact numerical correspondence of the two methods, since they are not identical in specific modeling and normalizing. Additionally, the definition of CD efficiency in quasilinear simulation is dimensional, whereas the linear model focuses on dimensionless efficiency parameters. There is also a distinction between the two models that has to be elucidated: in the linear model, we assume the LHW first modifies the distribution function, followed by ECW acting on this background distribution, from which the CD efficiency is calculated, but the quasilinear simulation is a full-wave code without any assumption like the linear model.

In quasilinear simulation, the power deposition is associated with the diffusion coefficient of a wave~\cite{shoucri1987application,SHOUCRI1989253}. Thus, the diffusion coefficient of ECW, the width of the parallel refractive index of ECW, and the diffusion coefficient of LHW are set to, respectively, 0.01, 0.01, and 0.015. This setting ensures that the RF powers are relatively small, and the ECW power is smaller than the LHW power. The ion temperature is $1.5$keV; the other parameters about plasma and magnetic field are the same as the linear case listed in Tab.\ref{tab: Plasma, magnetic field and LHW parameters in ECCD case}. We shift the LHW region to $3u_{e}-5u_{e}$ in this section. Because of the difference in normalization of the LHW diffusion coefficient between the two methods, the constant $D_{LH,0}$ in the linear model should be approximately three orders of magnitude greater than that in the quasilinear simulation. Thus, we set $D_{LH,0} \simeq 10$ in the linear model. This value poses a risk of invalidating the entire linear model based on the perturbation theory. But the linear model is still effective around some $y$ where the synergy effect begins to occur, since at this stage only a small fraction of the ECW resonance curve in velocity space couples to the perturbative response function; this comparison between the quasilinear simulation and the linear model still maintains validity as a reference.

%\cjn{It's notable that the diffusion coefficient of LHW in the quasilinear simulation is smaller $\nu_{e0} \sim 10^{3} \mathrm{s}^{-1}$ level than it in the linear mode, therefore the diffusion coefficient of LHW has to be approximately set to $D_{LH,0} \sim 15$ in corresponding linear model. This value poses a risk of invalidating the entire linear model based on the perturbation theory. But the linear model is still effective around some $y$ where the synergy effect begins to occur, since at this stage only a small fraction of the ECW resonance curve in velocity space couples to the perturbative response function. Therefore, this comparison between the quasilinear simulation and the linear model still maintains validity as a reference.}

\begin{figure}[htbp] 
	\centering
	\includegraphics[width=0.47\linewidth]{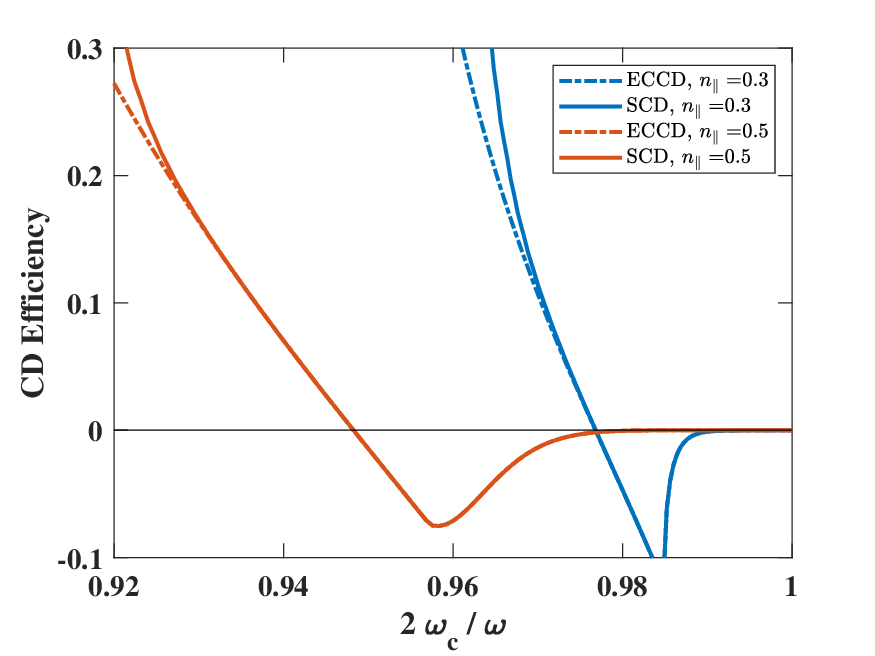}
	\includegraphics[width=0.47\linewidth]{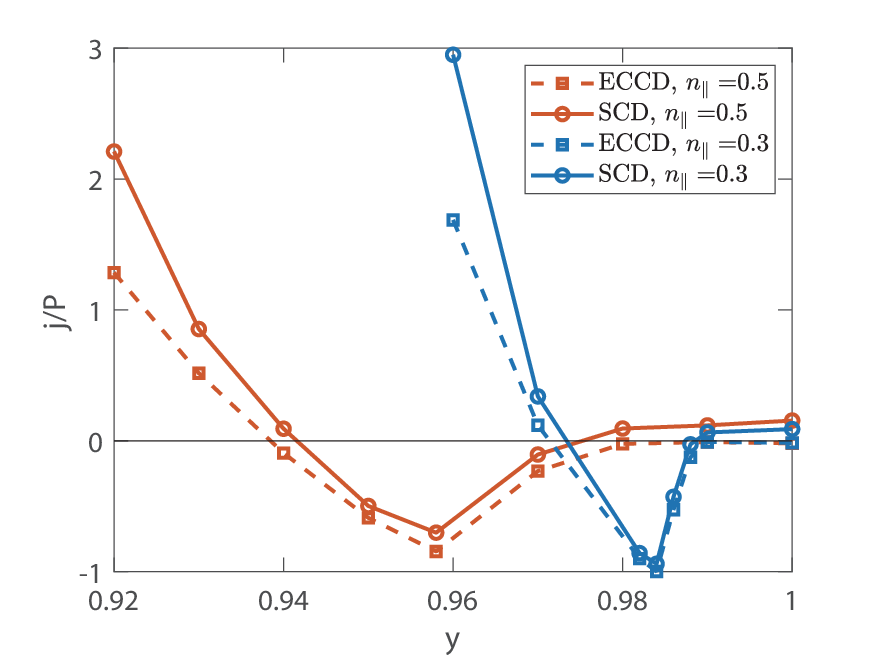}
	\caption{The comparison of the efficiency of SCD and ECCD between the linear model~(left panel) and the quasilinear simulation~(right panel).}
    \label{fig: comparison between linear model and quasilinear model}
\end{figure}

As shown in Fig.\ref{fig: comparison between linear model and quasilinear model}, ECCD and SCD efficiency as a function of $y$ with $n_{\parallel}=0.3$ and $0.5$ at $\theta_{p} = 0^{\circ}$ where located at the weak-field side of a flux is depicted by using linear model~(left panel) and quasilinear simulation~(right panel). The data shows that the results of the quasilinear simulation are consistent with those of the linear case in trends, both of which show the existence of the synergy effect at smaller $y$ for each $n_{\parallel}$ case.

However, the quasilinear simulation reveals a key feature absent in linear analysis: for each $n_{\parallel}$ case, SCD efficiency consistently exceeds ECCD efficiency at relatively larger $y$. Understanding this difference between linear and quasilinear models is valuable to clarify the limitations of the linear model. In essence, this difference suggests a nonlinear enhancement mechanism, which primarily originates from the kinetic process of high-energy electrons that are not accounted for in the linear model. To elucidate the mechanism underlying this CD efficiency enhancement, we choose wave parameters as $y = 0.99$ and $n_{\parallel} = 0.5$ to depict the electron distribution after reaching steady-state, the resonance curve of ECW, and the diffusion of LHW where fixed at $3u_{e}$ to $5u_{e}$, in velocity space.

\begin{figure}[htbp] 
	\centering
	\includegraphics[width=0.7\linewidth]{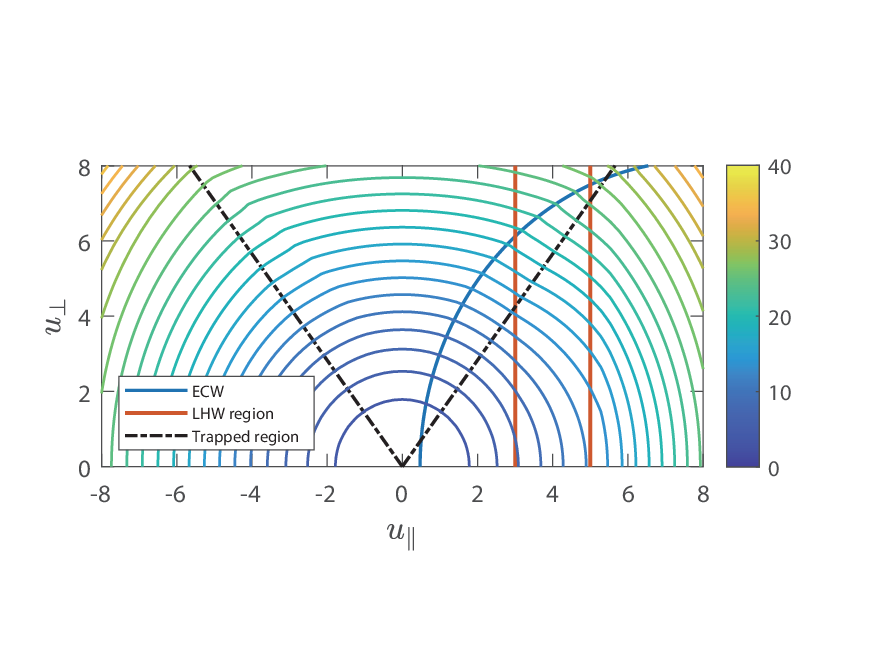}
	\caption{The electron distribution function in velocity space.}
        \label{fig: the electron distribution function in velocity space}
\end{figure}

As shown in Fig.\ref{fig: the electron distribution function in velocity space}, the Maxwellian distribution located at the diffusion region of LHW (the area between the two vertical orange lines) is modified in the parallel velocity direction, the distribution of ECW passing through (the blue line) is modified in vertical direction, and the portion altered by ECW in the trapped region is symmetric due to the pitch-angle scattering.  Those high-energy electrons in the trapped region pushed by ECW will go through the slow-downing process, although the probability of electrons escaping from both sides of the trapped region is equal, the electrons escaping on the right side can be further accelerated by LHW, resulting in asymmetric electrons relaxation in velocity space, and it causes the appearance of additional driving current, thus improving SCD efficiency. 

\begin{figure}[htbp] \label{fig: the CD situation with different diffusion coefficients}
	\centering
	\subfigure{
		\includegraphics[width=0.47\linewidth]{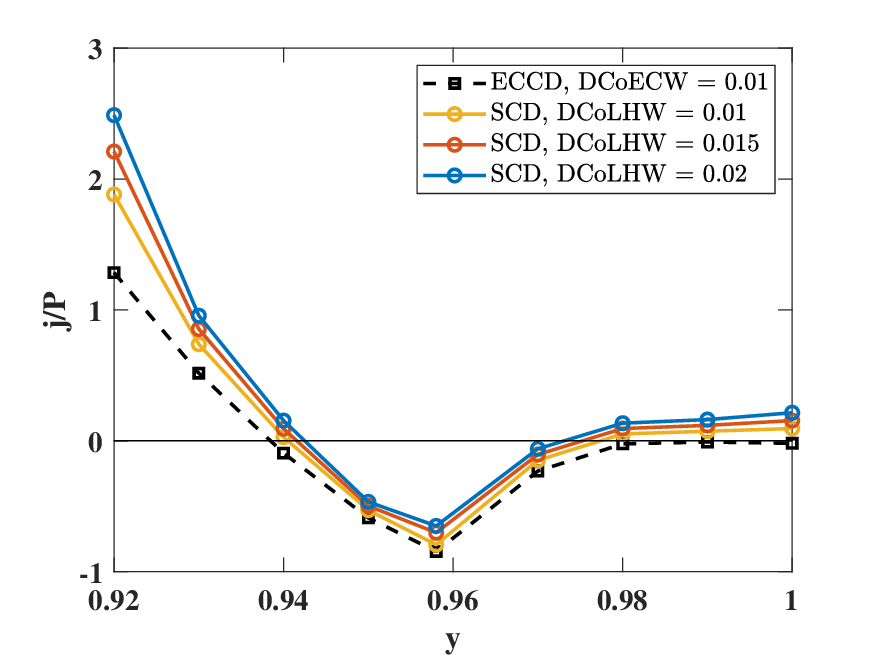}
	}
	\subfigure{
		\includegraphics[width=0.47\linewidth]{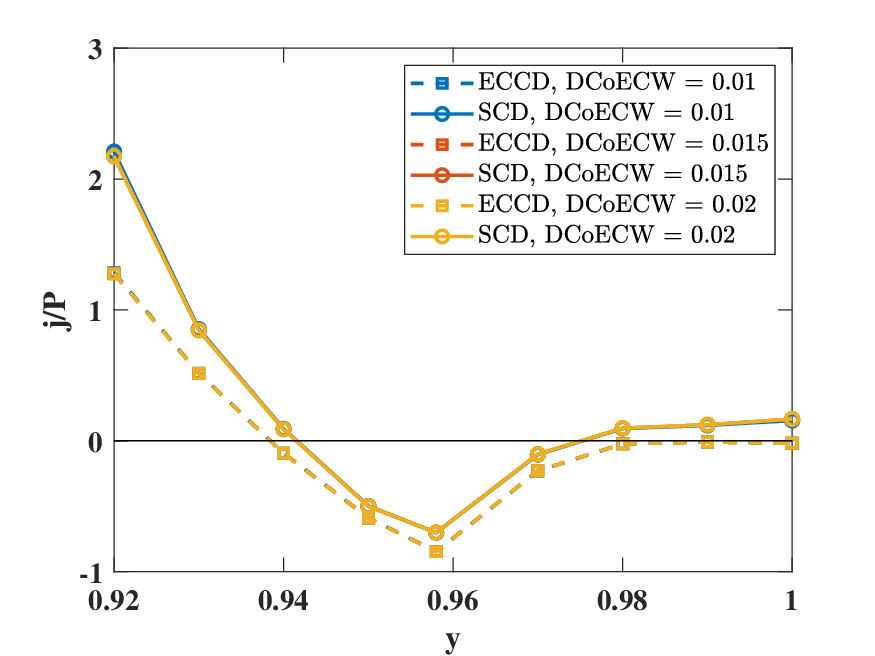}
	}
	\caption{The CD situations with different diffusion coefficients at $\theta_{p} = 0$ and $n_{\parallel} = 0.5$. (a)DCoLHW and (b)DCoECW mean, respectively, the diffusion coefficient of LHW and ECW.}
\end{figure}

In addition, the asymmetric electrons' relaxation depends on the LHW power instead of the ECW power. We gradually change the diffusion coefficients of LHW and ECW with the amplitude of 0.005 in the $n_{\parallel}=0.5, \theta=0^{\degree}$ case to probe their impact on SCD in quasilinear simulation. As shown in Fig.\ref{fig: the CD situation with different diffusion coefficients}(a), the data indicates that the SCD efficiency is gradually stronger than the ECCD case at smaller $y$ as the diffusion coefficient of LHW increases, which is consistent with the linear case manifested in Fig.\ref{fig: The SCD efficiency with different D_{LH}}. The enhancement of SCD efficiency compared to the ECCD case at larger $y$ remains, and it also depends on the LHW power. Fig.\ref{fig: the CD situation with different diffusion coefficients}(b) shows that both ECCD and SCD efficiency are independent of the ECW power, which is consistent with the linear analysis.

An additional minor factor is that LHW induces enhanced ECW power deposition in quasilinear simulations. However, when computing SCD efficiency using Eq.(\ref{eq: the definition of SCD in quasilinear simulation}), we retain the ECW power $P_{ECW}$ in Eq.(\ref{eq: thee definition of ECCD in quasilinear simulation}), which consequently leads to slightly overestimated $\xi_{SCD}$ values.

% In the presence of LHW, the ECW power deposition is increased slightly and additional driving current is generated, resulting in a larger $j_{LHW+ECW}$ term in Eq.(\ref{eq: the definition of SCD in quasilinear simulation}). Since the ECW power $P_{ECW}$ adopted in Eq.(\ref{eq: the definition of SCD in quasilinear simulation}) is the same as that in Eq.(\ref{eq: the definition of ECCD in quasilinear simulation}), $\xi_{SCD}$ is slightly larger than $\xi_{ECCD}$ except for smaller $y$. 

% \noindent 
% There are two main reasons for this disagreement. One is that LHW power leads to an increase in ECW power deposition. 
% Since the power deposition is associated with the diffusion coefficients in the quasilinear simulation,

To sum up, the additional enhancement of CD efficiency manifested by the quasilinear simulations at smaller $y$ for each $n_{\parallel}$ case, which is different from the linear case, arises primarily from kinetic processes involving high-energy electrons, with minor contributions from computational methodology differences.

% 另外，我们发现，双波协同的机制导致LHCD的效率没有提升，这一点明确地在以前地文章中例如 PRL中有提到。从定义、机制等等角度解释一下。

\subsection{SCD with parameters used in current reactors} \label{sec: the SCD results with parameters used in current reactors}

For benchmarking with the ECCD model~\cite{LinLiu2003}, the parameters in the linear model have been chosen as the same as those in it, as well as the X2-mode of ECW, to calculate ECCD and SCD efficiency, and show relevant results in previous sections. 

Given considerable significant developments in present reactors, the parameters should be improved to the level of current mainstream Tokamaks, and the ECW mode should be changed to O1-mode; only thus can we demonstrate the applicability and contribution of this linear model.

For the O1-mode, the $D_{ECW}$ can be approximated by 
    \begin{eqnarray}
        D_{O1} = \pi e^{2} |E_{\parallel}|^{2} \left( \frac{k_{\perp} u_{\perp}}{ 2 \omega_{c} } \right)^{2\ell} \frac{u_{\parallel}^{2}}{\gamma^{2}} \, ,
    \end{eqnarray}
    with $\ell=1$.
    Thus, the formulation of CD efficiency $\zeta^{\star}_{O1}$ with the O1-mode of ECW becomes
    \begin{eqnarray}
        \zeta^{*}_{SCD,O1} = - \frac{4}{\ln\Lambda} \langle \frac{B}{B_{max}} \rangle \frac{m u_{e}^{2} \int \mathrm{d} \gamma u_{\perp}^{2} \left[  c^{2}( \gamma^2 - 1 ) - u_{\perp}^{2} \right] f_{M} \tilde{\Lambda} \tilde{\chi}  }{\int \mathrm{d} \gamma u_{\perp}^{2} \left[  c^{2}( \gamma^2 - 1 ) - u_{\perp}^{2} \right] f_{M}  } \, .
    \end{eqnarray}
  
Three sets of plasma parameters~\cite{Poli2018, Zhuang_2019} corresponding to mainstream reactors ITER, DEMO1, and CFETR, as shown in Tab.\ref{tab: nowadays reactors}, are chosen to calculate ECCD and SCD efficiency. The other parameters are fixed with $n_{\parallel} = 0.7$, which corresponds the torodial lanch angle $20^{\degree}$, $D_{LH,0} = 0.2$, $\varepsilon = 0.2$, and the LHW region $2u_{e}-4u_{e}$.
    \begin{table}[htbp]
        \centering
        \begin{tabular}{c c c c c c c c}
        \hline
             Case & Reactor & $R_{0}$(cm) & $a$(cm) & $B_{0}$(T) & $n_{e0}$ ($10^{19} \mathrm{m}^{-3}$) & $T_{e0}$(keV) & $Z_{\mathrm{eff}}$  \\
             \hline
             (a) & ITER & 620 & 201 & 5.3 & 10.56 & 24.49 & 1.76  \\
             (b) & DEMO1 & 907 & 292 & 5.66 & 10.50 & 33.25 & 2.0\\
             (c) & CFETR(A.4) & 772 & 220 & 6.5 & 13.1 & 32 & 2.45 \\ 
        \hline
        \end{tabular}
        \caption{The parameters used in current reactors.}
        \label{tab: nowadays reactors}
    \end{table}
    \begin{figure}[htbp]
        \centering
        \includegraphics[width=0.3\textwidth]{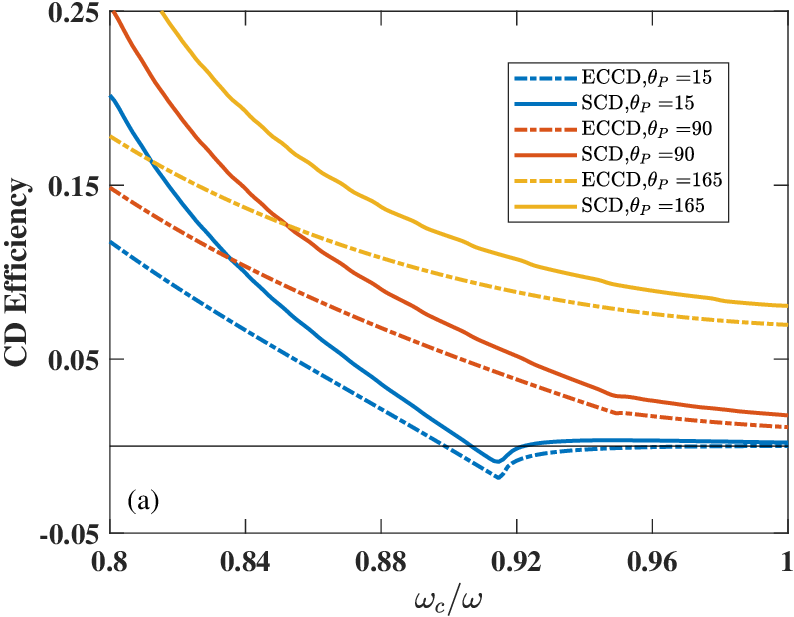}
        \includegraphics[width = 0.3\textwidth]{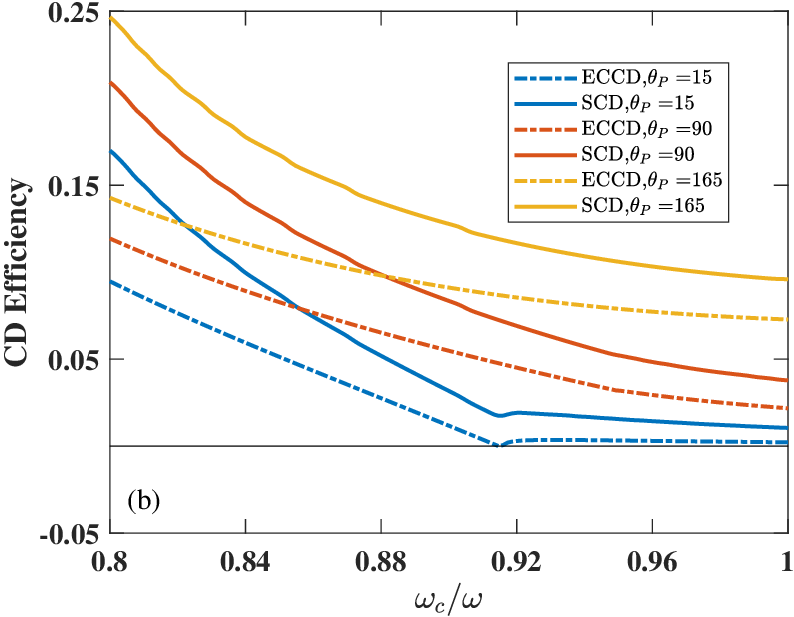}
        \includegraphics[width = 0.3\textwidth]{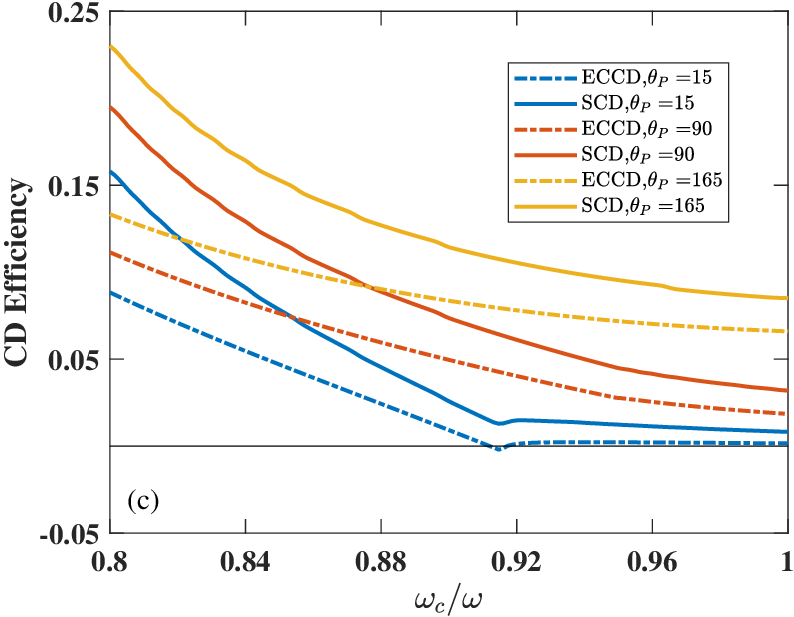}
         \caption{The ECCD~(dashed line) and SCD~(solid line) efficiency with parameters of ITER~(a), DEMO~(b), CFETR(A.4)~(c).}
        \label{fig: results of nowadays reactors}
    \end{figure}

As Fig.\ref{fig: results of nowadays reactors} shows, all three cases demonstrate the presence of the synergy effect when $y=\omega_{c}/\omega$ is relatively small, with the synergy effect being most pronounced under ITER parameters. By comparing several panels, we can discover the influence of parameters on the synergy effect. When the density and $Z_{eff}$ differ only slightly, as indicated by cases (a) and (b), an increase in temperature suppresses the ECCD and SCD efficiency. In contrast, when the temperature difference is minimal, as indicated by cases (b) and (c), the efficiencies of ECCD and SCD are very similar, implying that other parameters have a marginal influence on CD efficiency. The results reveal the strong sensitivity of ECCD to temperature, and SCD can be a tool to destabilize the sensitivity of ECCD to temperature.

\subsection{Integrating SCD into ray-tracing code}

The linear model is not only beneficial for elucidating the underlying mechanisms of SCD and quantifying its sensitivity to variations in key parameters such as density, temperature, and wave frequency, but also shows the potential possibility to integrate it into ray-tracing codes for fast calculating SCD scheme and supporting experimental works as references. 

Ray-tracing codes are widely used in fusion plasma to model waves' launch, propagation, absorption, and CD. The entire process of ECW interacting with plasma under IETR conditions had been benchmarked with many ray-tracing codes~\cite {Prater_2008}. We choose GENRAY~\cite{genray_smirnov1994general}, one of the widely used ray-tracing codes, to quantify the synergy effect and show the potential possibility in this section. The code employs the cold-plasma dispersion relation in combination with the ray-tracing equations to calculate the wave propagation. The wave power deposition can be obtained by calculating the damping rate of the wave in the plasma and integrating it along the wave trace. The current driven by ECW can be computed by the ECCD model~\cite{LinLiu2003}. The essential step is that we use the output data of GENRAY as the input data of the linear model to calculate the current profiles of ECCD and SCD.

\begin{figure}[htbp]
    % \centering
    \includegraphics[width = 0.3\linewidth]{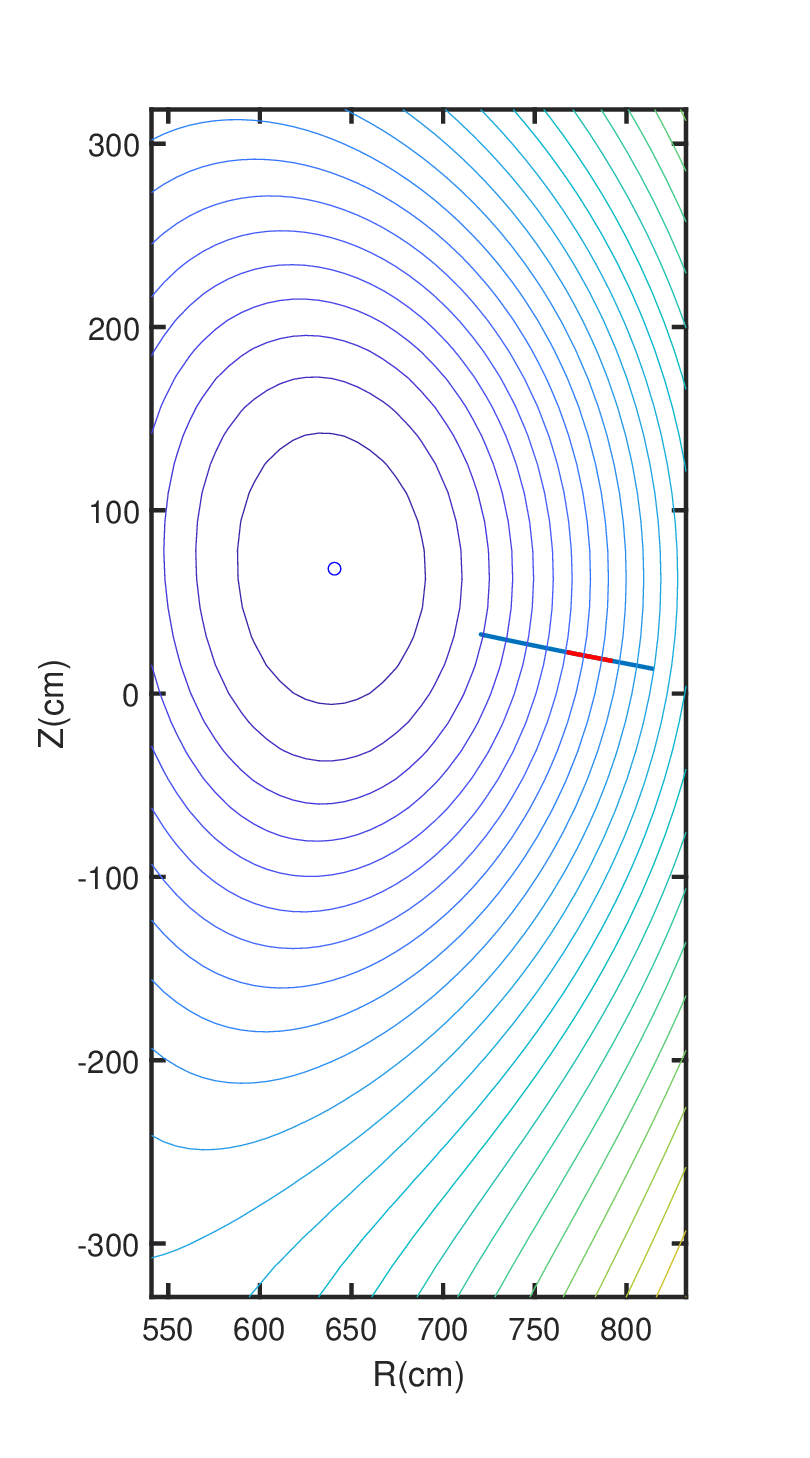}
    \hspace{-1.0cm}
     \raisebox{+0.5mm}{
    \includegraphics[width=12.8cm, height=8.3cm]{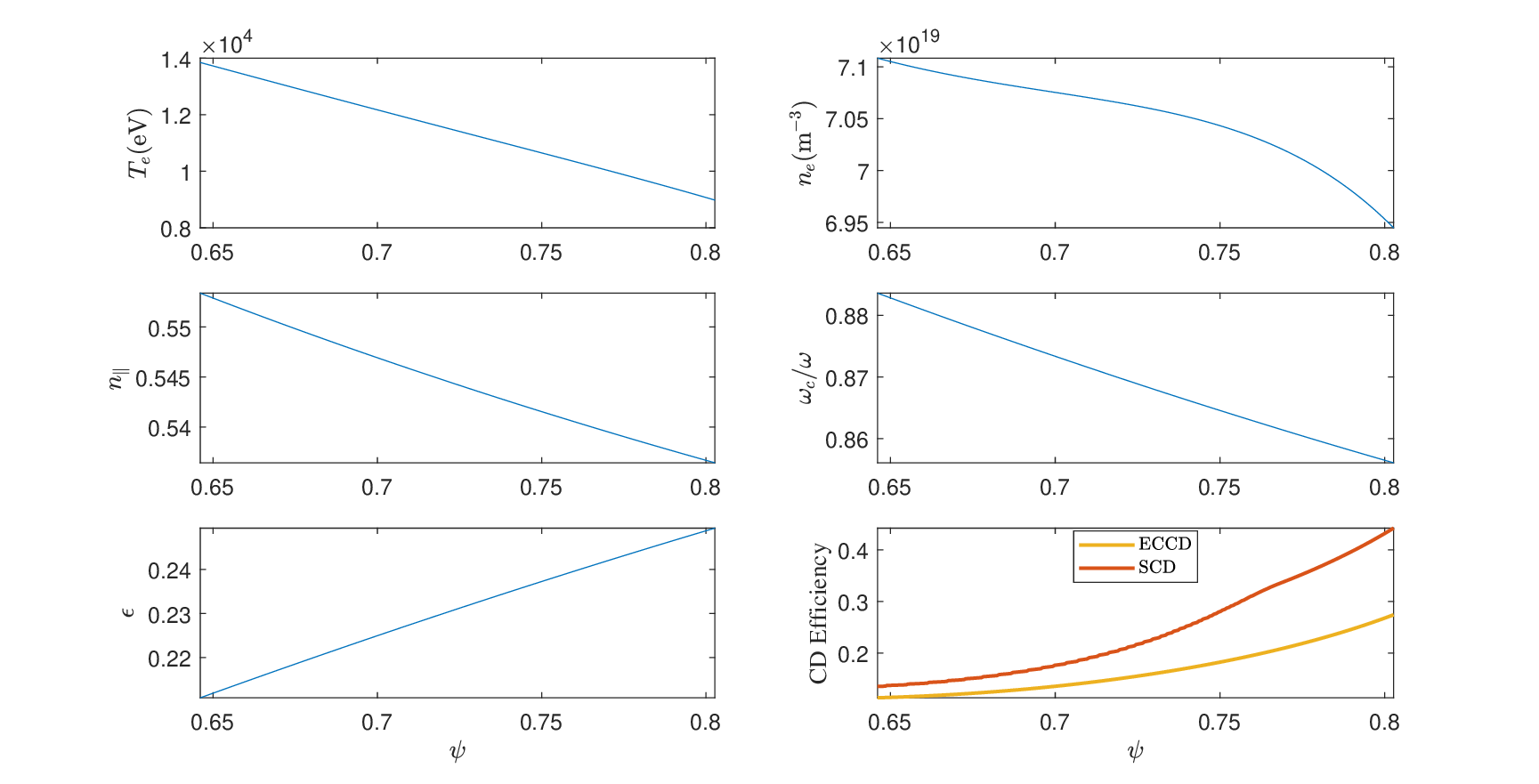}
    }
    \caption{Left panel: the entire ray trace of ECW in plasma marked by the blue line, the power deposition part marked by the red line. The ITER configuration is used here; Right panel:  The parameters used in the linear model, which come from the output data of GENRAY, vary with flux $\psi$, where the power deposition. The ECCD~(orange line) and SCD~(red line) efficiency calculated by the linear model is provided in the last subgraph.}
    \label{fig: wave trace}
\end{figure}

The ITER configuration of plasma and magnetic field is adopted here. The initial poloidal and toroidal launch angles of ECW are set to $10$ and $210$ degrees, respectively, and the initial position of the cone was set to $(8.8\mathrm{m}, 0\mathrm{m})$. The frequency of the ECW is set to $170$GHz, and the injection of ECW power equals $1$MW.  The left panel of Fig.\ref{fig: wave trace} describes the wave trace of ECW indicated by the blue line, and the location of the power deposition is marked by the red line.

\begin{figure}[htbp]
    \centering
    \includegraphics[width=0.45\linewidth]{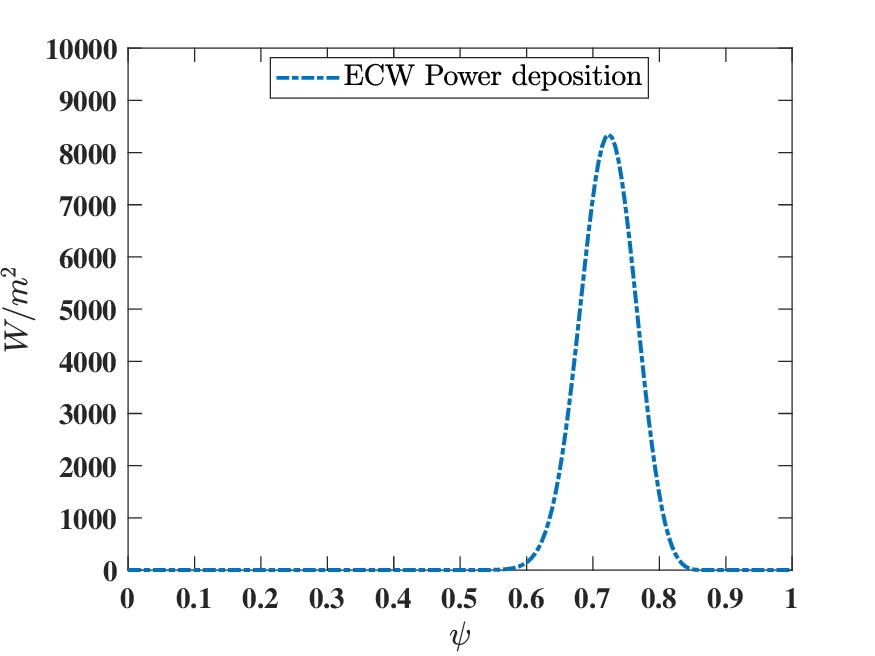}
    \includegraphics[width=0.45\linewidth]{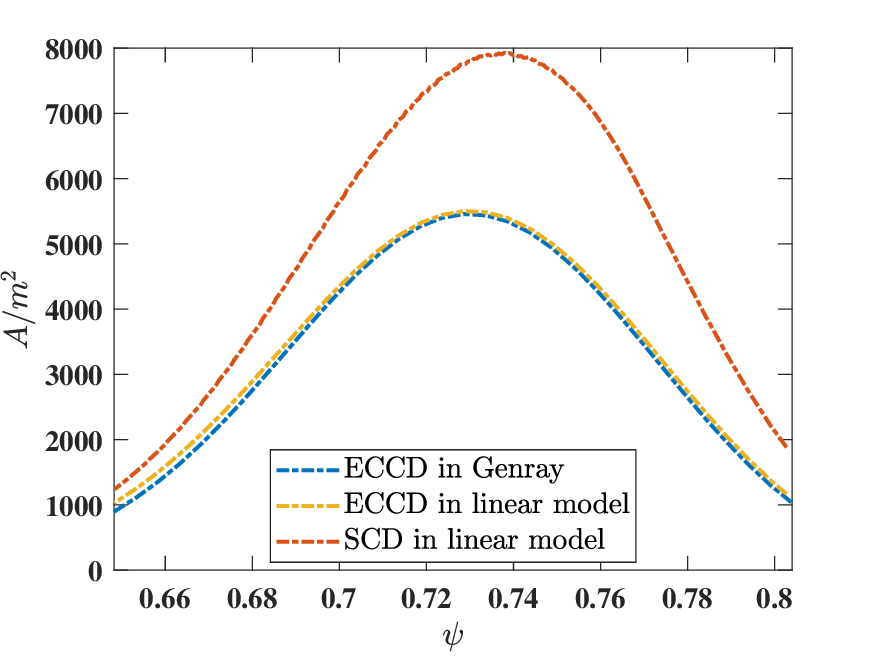}
    \caption{Left panel: the power deposition of ECW varying with flux $\psi$. Right panel: the ECCD~(orange dotted line) and SCD~(red dotted line) profiles calculated by the linear model, and the ECCD profiles~(blue dotted line) resulted from GENRAY. }
    \label{fig: current profile}
\end{figure}

The right panel of Fig.\ref{fig: wave trace} shows the parameters $T_{e}$, $n_{e}$, $n_{\parallel}$, $y=\omega_{c}/\omega$ and $\epsilon$ varying with the flux $\psi$ during the wave deposition. These parameters are originally provided by the output data of GENRAY; we translate the data to parameters used in the linear model according to basic formulas, for example, $y$ is obtained by $y=\ell \omega_{c}/\omega$. As for the LHW effect, we assume that its power deposition has occurred near the ECW deposition region. Considering the broad deposition profile and accessibility of LHCD, such a deposition setup can be readily achieved in off-axis regions. Further, $D_{LH,0}$ is set to $0.2$, and the LHW region is set to $2u_{e} - 4u_{e}$. Using these parameters can compute the ECCD and SCD efficiency, as shown in the last subgraph of the right panel of Fig.\ref{fig: wave trace}. Then, combining with the data of ECW power deposition provided by GENRAY, as shown in the left panel of Fig.\ref{fig: current profile}, the current profile can be obtained. 

The right panel of Fig.\ref{fig: current profile} shows the agreement between GENRAY's ECCD profile~(blue dotted line) and the linear model's ECCD profile~(orange dotted line), since the same theory~\cite{LinLiu2003} is used here. SCD profile~(red dotted line) has a considerable improvement compared to the ECCD profile. The total current of ECCD is approximately equal to $17.1$kA, and the total current of SCD is approximately equal to $24.7$kA. This suggests the potential application: integrating this linear model into ray-tracing code, such as GENRAY, enables fast evaluation of SCD that reduces computational cost compared to full-wave codes, and provides a reference for experimental works. 

\section{The criteria of the synergy effect} \label{sec: criteria}
The advantage of a linear model compared to simulations and experiments is that we are able to probe the criterion of the occurrence of the synergy effect. The synergy efficiency according to Eq.(\ref{eq: linearized CD efficiency}) can be written as
\begin{equation} \label{eq: the synergy efficiency}
	\zeta_{syn}^{*} = - \frac{4}{\ln \Lambda} \langle \frac{B}{B_{m}} \rangle \frac{m u_{e}^{2} \int d\gamma (u_{\perp})^{2\ell} f_{M} \tilde{\Lambda} \tilde{\delta\chi}}{\int d\gamma (u_{\perp})^{2\ell} f_{M}},
\end{equation}
\noindent Obviously, if the integral in the numerator is in the case
\begin{equation} \label{eq: criterion of the synergy effect}
	mu_{e}^{2} \int (u_{\perp})^{2\ell} f_{M} \tilde{\Lambda} \tilde{ \delta \chi }  > 0,
\end{equation}
the synergy effect will occur. Here, the lower and upper bound of integral are $max\left[ \gamma(u_{\parallel,1}), \gamma_{min}  \right]$ and $min\left[ \gamma(u_{\parallel,2}), \gamma_{max}  \right]$, in which $\gamma(u_{\parallel,1})$ and $\gamma(u_{\parallel,2})$ represent $\gamma$ calculated by resonance relation $\delta( \gamma - \frac{n_{\parallel} u_{\parallel}}{c} - y  )$ with the left boundary $u_{\parallel,1}$ and the right $u_{\parallel,2}$ of LHW region, and $\gamma_{min}$, $\gamma_{max}$ are the lower and upper bounds determined by Eq.(\ref{eq: the bounds of ECW resonance}). The reason that Eq.(\ref{eq: criterion of the synergy effect}) can be considered as a criterion for the occurrence of the synergy effect would be explained by the matching pattern of ECW's resonance curves and LHW's diffusion region in velocity space.
\begin{figure}[htbp] 
	\centering
	\includegraphics[width=0.8\linewidth]{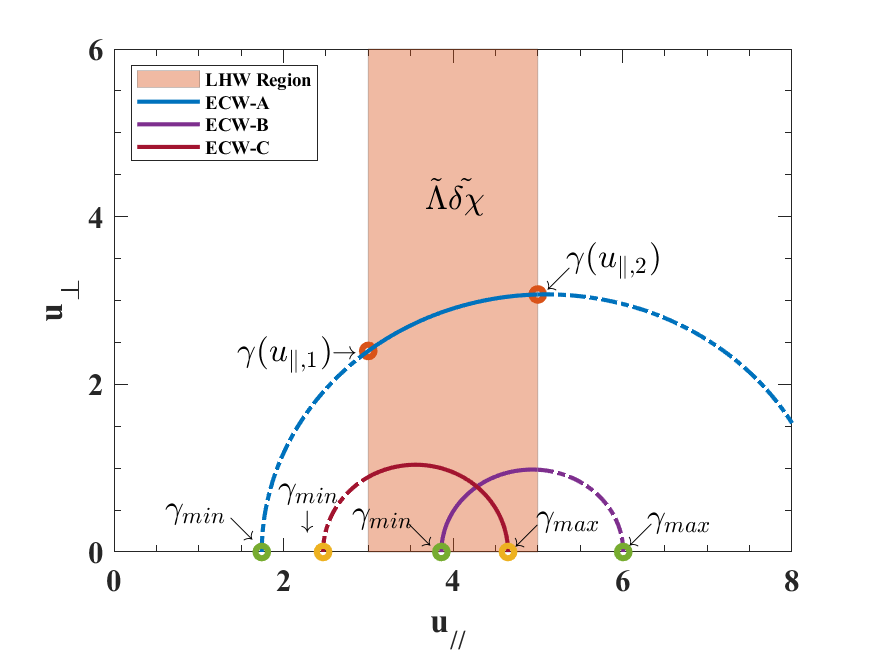}
	\caption{The resonance curves of ECW and the diffusion region of LHW in velocity space. The meaning of each point is marked in the figure. The different color lines denote different CD cases: ECW-A($y = 0.95, n_{\parallel} = 0.4$), ECW-B($y = 0.92, n_{\parallel} = 0.4$), and ECW-C($y = 0.958, n_{\parallel} = 0.3$). The LHW diffusion region is fixed at a range of $3u_{e}$ to $5u_{e}$.}
    \label{fig:  the three situations of ECW resonance curve}
\end{figure}

There are three typical situations in which resonance curves of ECW are located in velocity space without considering the trapped region, as shown in Fig.\ref{fig:  the three situations of ECW resonance curve}, we shift the LHW region used in above calculations to $3u_{e}-5u_{e}$ for clear drawing and explanation. Each part in the criterion Eq.(\ref{eq: criterion of the synergy effect}) has its physical meaning in velocity space. The term $(u_{\perp})^{2\ell}$ originates from the requirement of the X-mode of ECW, which is replaced by $u_{\perp}^{2} \left[  c^{2}( \gamma^2 - 1 ) - u_{\perp}^{2} \right]$ in O1-mode of ECW. The Maxwellian distribution $f_{M}$ represents that most electrons exist in the low-energy area instead of the high-energy one. The perturbation response function $\tilde{ \delta \chi }$ describes additional electron relaxation caused by LHW in a given region, but $mu_{e}^2\tilde{\Lambda} \tilde{ \delta \chi }$ wouldn't appear outside of the given region. Thus, the dashed line part of the resonance curve of ECW has no integral contribution to the synergy effect, and only the solid line part of the resonance curve can take effect. Correspondingly, the lower and upper bounds of integral change to $max\left[ \gamma(u_{\parallel,1}), \gamma_{min}  \right]$ and $min\left[ \gamma(u_{\parallel,2}), \gamma_{max}  \right]$ as illustrated in Tab.\ref{tab: the bounds of the integral in different cases.}.
\begin{table}[htbp] 
	\centering
	\caption{The bounds of the integral in different cases.}
	\begin{tabular}{c c c}
		\hline
		Case & lower bound & upper bound \\
		\hline
		ECW-A & $\gamma(u_{\parallel,1})$ & $\gamma(u_{\parallel,2})$ \\
		ECW-B & $\gamma_{min}$ & $\gamma(u_{\parallel,2})$ \\
		ECW-C & $\gamma(u_{\parallel,1})$ & $\gamma_{max}$ \\
		\hline
	\end{tabular}
        \label{tab: the bounds of the integral in different cases.}
\end{table}

The reason that we neglect the trapped effect is that the synergy effect tends to occur directly in the low-energy part of the LHW diffusion region, where a large number of electrons exist. If the trapped boundary intersects with the resonance curves of ECW within the LHW diffusion domain, the intersection must occur in the high-energy region, and the resulting modification only changes the left bound of the integral. Considering there are fewer electrons in the high-energy area, the remaining resonance curve of ECW makes a small contribution to the integral, even if the synergy effect occurs, it will be very inconspicuous. 

\begin{figure}[htbp]
	\centering
	\subfigure{
	\includegraphics[width=0.45\linewidth]{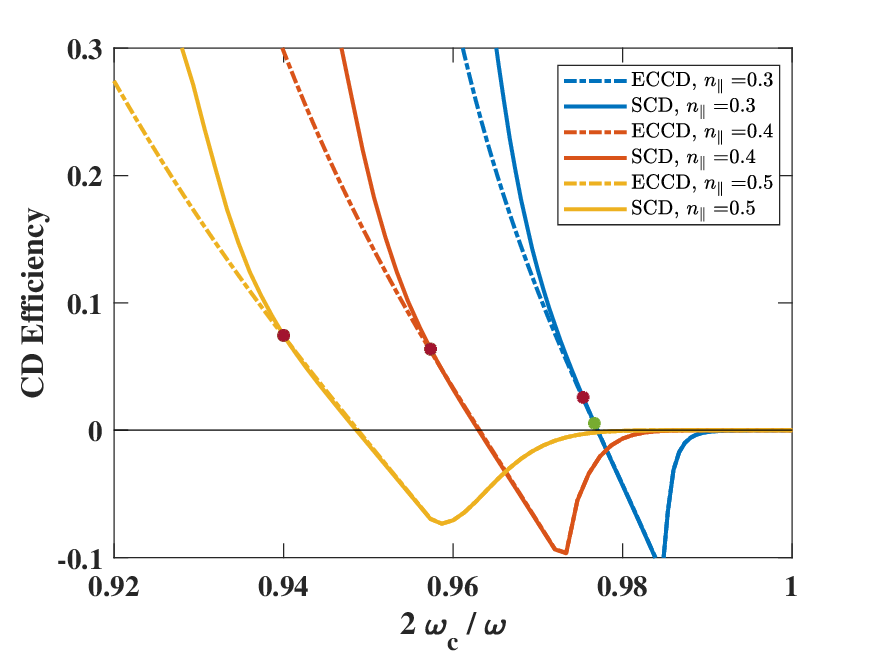}
	}
	\subfigure{
	\includegraphics[width=0.45\linewidth]{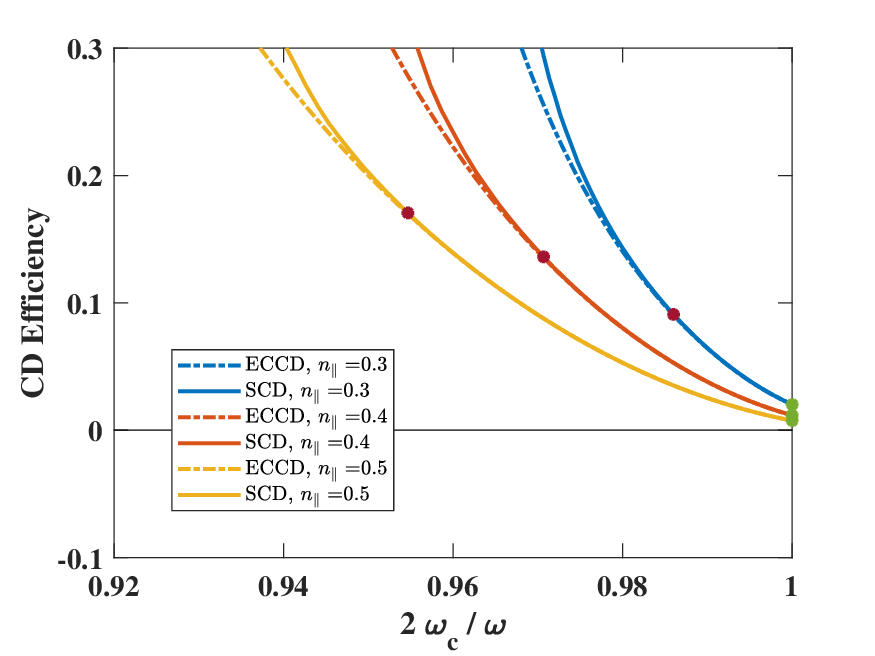}
	}
	\caption{The points of criteria are marked at different poloidal angles (a)$\theta_{p} = 15^{\circ}$ and (b)$\theta_{p} = 165^{\circ}$. The green and red points determined by Eq.(\ref{eq: criterion of the synergy effect}) and Eq.(\ref{eq: the synergy effect sufficiently important}), respectively, represent the threshold values of occurrence and sufficient significance of the synergy effect. }
    \label{fig: the points of criteria}
\end{figure}

Additionally, although the criterion is established by Eq.(\ref{eq: criterion of the synergy effect}), the synergy effect still can be insignificant due to the larger denominator in Eq.(\ref{eq: the synergy efficiency}). We introduce two new definitions 
\begin{equation}
\fl	\kappa =  mu_{e}^2 \int_{max\left[ \gamma(u_{\parallel,1}), \gamma_{min}  \right]}^{min\left[ \gamma(u_{\parallel,2}), \gamma_{max}  \right]}  d\gamma (u_{\perp})^{2\ell} f_{M} \tilde{\Lambda} \tilde{\delta\chi} \bigg/ \int_{max\left[ \gamma(u_{\parallel,1}), \gamma_{min}  \right]}^{min\left[ \gamma(u_{\parallel,2}), \gamma_{max}  \right]} d\gamma (u_{\perp})^{2\ell} f_{M} ,
\end{equation}
and
\begin{equation}
\fl	\eta = \int_{ max\left[ \gamma(u_{\parallel,1}), \gamma_{min}  \right] }^{ min\left[ \gamma(u_{\parallel,2}), \gamma_{max}  \right] }  d\gamma (u_{\perp})^{2\ell} f_{M} \bigg/ \int_{\gamma_{min}}^{\gamma_{max}} d\gamma (u_{\perp})^{2\ell} f_{M},
\end{equation}
where $\eta$ is the power factor that represents the ECW power deposited in the diffusion region of LHW divided by the total ECW power deposition, i.e., the integral represented by the solid part of the resonance curve of ECW divided by the integral represented by the total (solid + dashed part) resonance curve of ECW, and $\kappa$ is the CD efficiency of the ECW power deposited in the diffusion region of LHW. Therefore, the synergy efficiency reads
\begin{equation}
	\zeta^{*}_{syn} = - \frac{4}{\ln \Lambda} \langle \frac{B}{B_{m}} \rangle \kappa \eta.
\end{equation}
Obviously, the final magnitude of the synergy effect is determined by $\kappa$ and $\eta$ together at a given flux surface.   Actually, $\eta$ quantifies the degree of the overlap of the two waves' quasilinear domains. We consider the criterion that the synergy effect is sufficiently significant in the numerical calculation as 
\begin{equation}\label{eq: the synergy effect sufficiently important}
	\eta > 1\times10^{-4}.
\end{equation} 
We depict $\tilde{\zeta}_{syn} $ varying with $y$ and mark threshold values of the two criteria as shown in Fig.\ref{fig: the points of criteria}, the parameters are the same as in Tab.\ref{tab: Plasma, magnetic field and LHW parameters in ECCD case} except for $n_{\parallel} = 0.3, 0.4, 0.5$. 
% The specific threshold values of $y$ corresponding to the marked points in Fig.\ref{fig: the points of criteria} are shown in Tab.\ref{tab: the specific threshold values.}.
% \begin{table}[htbp]
% 	\centering
% 	\caption{The specific threshold values of occurrence and sufficient significance of the synergy effect.}
% 	\begin{tabular}{c c c c }
% 		\hline
% 		$\theta_{p} = 15^{\circ}$ & $n_{\parallel}=0.3$ & $n_{\parallel}=0.4$ & $n_{\parallel}=0.5$  \\
% 		\hline
% 		Point &  y  & y  & y  \\
% 		\hline
% 		Green & 0.9887     & 0.9627       & 0.9373   \\
% 		Red & 0.9760       & 0.9547     & 0.9307  \\
% 		\hline
% 	\end{tabular}
% 	\\[5pt]
% 	\begin{tabular}{c c c c}
% 		\hline
% 		$\theta_{p} = 165^{\circ}$ & $n_{\parallel}=0.3$ & $n_{\parallel}=0.4$ & $n_{\parallel}=0.5$  \\
% 		\hline
% 		Point &  y &  y  & y  \\
% 		\hline
% 		Green & 1  & 1   & 1   \\
% 		Red & 0.9800 & 0.9587 & 0.9373  \\
% 		\hline
% 	\end{tabular}
%         \label{tab: the specific threshold values.}
% \end{table}

The above data manifest that the strong-field side of a flux surface has a broader $y$ scope for the synergy effect, and it is significant enough only when the power factor is larger than $0.01\%$. The reason that the synergy effect is inclined to take place at the strong-field side of a given flux surface instead of the weak-field side is that the trapped region of the former is smaller, and there are relatively more passing electrons.  In general, the two criteria imply that the synergy effect quantitatively depends on the power factor $\eta$, the constant $D_{LH}$ determined by the LHW power, and the number of synergy electrons.

Optimizing the synergy effect requires improvements in the above three aspects by combining practical experimental conditions.
The power factor $\eta$ is a function of several variables 
% including $y$, $n_{\parallel}$, $u_{\parallel,1}$, $u_{\parallel,2}$, and $T_{e}$. 
% Among these variables, 
$y$ and $n_{\parallel}$ influence the shape of the resonance curve of ECW, while $u_{\parallel,1}$ and $u_{\parallel,2}$, determined by Landau damping, extract the relevant portion contributing to the synergy effect. It is worth noting that the integral $\int_{\gamma_{min}}^{\gamma_{max}} d\gamma (u_{\perp})^{2\ell} f_{M}$ is non-linear along the resonance curve due to the exponential decrease of $f_{M}$ as $\gamma$ increases. Consequently, the low-energy part (i.e., the smaller $\gamma$) of the resonance curve contributes more significantly compared to the high-energy part (i.e., the larger $\gamma$). Therefore, optimizing $\eta$ involves maximizing the involvement of the low-energy portion of the resonance curve of ECW within the diffusion region of LHW by appropriately adjusting relevant parameters. Besides, the constant $D_{LH}$, determined by the LHW power, directly affects the magnitude of $\tilde{\delta \chi}$ that represents the degree of electron relaxation induced by LHW power in velocity space. Hence, increasing the LHW power within the limitations of the linear model is beneficial for enhancing the synergy effect. Finally, the synergy electrons, which are accelerated by two waves, serve as the primary contributors to the synergy effect. The function $\tilde{\delta \chi}$ describes the electrons' response under the waves' effect, but the term $f_{M}$ in the integral of $\kappa$ decides the number of synergy electrons in velocity space. Thus, more synergy electrons will make the synergy effect perform strongly.

Although we can not use quasilinear simulation to cross-validate the two criteria due to the non-linear nature of simulation, both criteria still show great application potential for quickly determining whether the synergy effect occurs and whether it's significant enough for given parameters. 
	
\section{Summary} \label{sec: summary}

% 双波协同除了提高效率有什么其他的优点：**弱化ECCD对于温度和密度敏感程度，用于离轴的电流剖面控制**

% 这篇文章的重点不在于调整参数寻求优化电流，而是一个基础理论

A linear model of SCD with LHW and ECW has been developed in this paper, offering a quantitative description of SCD compared to the previous qualitative understanding~\cite{Dumont2004theory}. The response function dominated by collisions in the presence of LHW is derived from the adjoint equations by using perturbation and Green-Function techniques, where the relativistic and trapping effects are considered. The SCD efficiency compared to the ECCD efficiency is given based on plasma parameters presented in~\cite{LinLiu2003}. The comparison results show that the synergy effect is inclined to occur at smaller $y=2\omega_{c}/\omega$ with the fixed $n_{\parallel}$ and $\theta_{p}$. By neglecting the trapping effect (via the limit $\epsilon \to 0$, the CD mechanism shifts predominantly to the Fisch-Boozer mechanism, and the SCD efficiency increases compared to cases with the trapping effect. The predictions of the linear model align well with trends observed in 2D quasilinear Fokker-Planck simulations, particularly the emergence of synergy at lower $y$. However, simulations reveal additional nonlinear enhancements, especially from the kinetic progress of high-energy electrons, which are not captured by the linear model. Given considerable development of current reactors, SCD efficiencies are calculated by using parameters of ITER, DEMO1, and CFETR, demonstrating the clear synergy effect in all cases, especially pronounced in the ITER configuration. Using output data from GENRAY as input data for the linear model to calculate the ECCD and SCD current profiles shows the potential application of integrating SCD into ray-tracing code. Two criteria for the occurrence and the sufficient significance of the synergy effect are proposed, which correspond to Eq.(\ref{eq: criterion of the synergy effect}) and Eq.(\ref{eq: the synergy effect sufficiently important}), respectively. The threshold value $y$ at which the synergy effect becomes significant shifts higher as $n_{\parallel}$ decreases. The criteria indicate that the synergy effect is dependent on the power factor $\eta$ that quantifies the degree of the overlap of two waves' diffusion domain, the LHW power $D_{LH,0}$, and the number of synergy electrons.

The most promising prospect of SCD in experiment is applying for off-axis current profile control~\cite{Dumont2000combined, PhysRevLett.93.255002}, since this scheme can be a tool to remove the strong sensitivity of ECCD to temperature and trapping. Finally, it should be mentioned that this linear model is universally applicable to other relevant parameters, not only for parameters that appeared in this study. The theoretical formulation of the above calculations also displays the possibility of generalizing the application to other synergetic CD based on the matching pattern of quasilinear domains in velocity space.

\ack
The authors would like to thank Professor Y.X. Long for providing us with the two-dimensional Fokker-Planck code. One of the authors, named J.N. Chen, would like to thank Professor R.J. Dumont for his positive regard on this work, and a fellow student named H. Fan for his help on programming. This work is supported by the National MCF Energy R\&D Program of China (Grant Nos.2019YFE03090400 and 2019YFE03030004), and the National Natural Science Foundation of China (Grant Nos.12375222 and 11775154).

\appendix

\section[]{The formulations of Lin-Liu's response function} \label{apd: Details of Lin-Liu's response function}
The zero-order response function given by Lin-Liu is 
\begin{equation}
	\tilde{\chi}_{0} = sgn(u_{\parallel}) F(u) H(\lambda),
\end{equation}
where
\begin{eqnarray}
	F(u) & =  \frac{1}{f_{c} u_{e}^{4}}\left(\frac{\gamma+1}{\gamma-1}\right)^{\hat{\rho} / 2} \int_{0}^{u} du^{\prime}\left(\frac{u^{\prime}}{\gamma^{\prime}}\right)^{3}\left(\frac{\gamma^{\prime}-1}{\gamma^{\prime}+1}\right)^{\hat{\rho} / 2} \cr
	& = \frac{1}{f_{c}}\left(\frac{u^{4}}{u_{e}^{4}}\right) \int_{0}^{1} d x x^{\hat{\rho}+3} \frac{1}{\left(1+(u x / c)^{2}\right)^{3 / 2}} \cr
	& \quad \times\left[\frac{1+\sqrt{1+(u / c)^{2}}}{1+\sqrt{1+(u x / c)^{2}}}\right]^{\hat{\rho}} ,
\end{eqnarray}
and 
\begin{equation}
	H(\lambda) \equiv \frac{\theta(1-\lambda)}{2} \int_{\lambda}^{1} \frac{d \lambda^{\prime}}{\left\langle\left(1-\lambda^{\prime} B / B_{m }\right)^{1 / 2}\right\rangle} .
\end{equation}
Here $\hat{\rho} \equiv \frac{Z_{\mathrm{eff}}+1}{f_{c}}$ and $f_{c}$ is the effective circulating particles fraction in the neoclassical transport theory, which is determined by  
\begin{eqnarray}
	f_{c} & \equiv \frac{3}{2}\left\langle\frac{B^{2}}{B_{m }^{2}}\right\rangle \int_{0}^{1} d \lambda H(\lambda) \\
	&=\frac{3}{4}\left\langle\frac{B^{2}}{B_{m }^{2}}\right\rangle \int_{0}^{1} \frac{\lambda d \lambda}{\left\langle\left(1-\lambda B / B_{m }\right)^{1 / 2}\right\rangle} .
\end{eqnarray}
% Additionally, it's notable that there is a typo in Eq.(A8) of \cite{LinLiu2003}, which should be 
% \begin{eqnarray}
%     \frac{\mathrm{d} H }{\mathrm{d} \lambda} = - \frac{1}{2} \frac{1}{( 1 - \lambda \langle h \rangle )^{1/2}} \frac{1}{ \left[ 1 + s^2( c_{2}(1-s^{2}) + c_{4}s^2 ) \right] }
% \end{eqnarray}

\section[]{The approximation of $F(u)$ } \label{apd: The approximation of Fu}
We suppose the relativistic $F(u)$, whose specific form is Eq.(A.2), and its derivatives can be approximated as simple functions with fixed order for the convenience of numerical calculation. In our calculations, we use \textit{mathematica} to fit $F(u)$, which is
\begin{equation}
	F(u) = (\frac{u}{u_{e}})^{3.1214}
\end{equation} 
with $C_{k} = 1$ and $k=3.1214$.
The numerical results of the two functions Eq.(A.2) and Eq.(B.1), and their derivatives are shown in Fig.B1. The blue, orange, and yellow lines denote, $F(u)$ itself, the first order derivative $\partial F(u) / \partial u$, and the second order derivative $\partial^{2}  F(u) / \partial u^2$, respectively. The solid line representing Eq.(B.1) is close enough to the dashed line representing Eq.(A.2). 
\begin{figure}[htbp]
	\centering
	\includegraphics[width=0.7\linewidth]{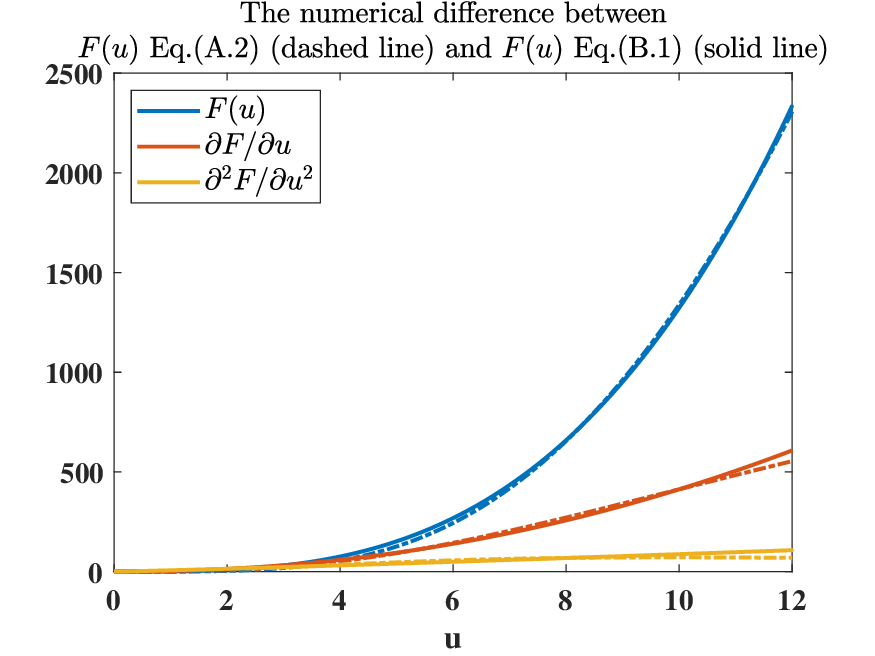}
	\caption{The numerical differences between Eq.(A.2)(dashed lines) and Eq.(B.1)(solid lines).}
\end{figure}
Then we calculate the statistical correlation between dashed and solid lines, as shown in Tab.B1.
\begin{table}[htbp]
	\centering
	\caption{Statistical correlation analysis between Eq.(A.2) and Eq.(B.1).}
	\begin{tabular}{c c c c}
		\hline
		 & Root mean square error & Correlation coefficient & P-value \\
		\hline
		$F(u)$ & 15.167617 & 0.999839 & 0~(approximated) \\
		$\partial F(u) / \partial u$ & 13.545415 & 0.997474 & 0~(approximated) \\
		$\partial^{2}  F(u) / \partial u^2$ & 12.370348 & 0.938761 & 0~(approximated)\\
		\hline
	\end{tabular}
\end{table}
The above data show that $F(u)$ itself and its first derivative $\partial F(u) / \partial u$ have strong statistical correlation between Eq.(A.2) and Eq.(B.1). Although the statistical correlation of second derivative $\partial^2 F(u) / \partial u^2$ is poor, considering the small contribution of higher-order derivative in differential equation Eq.(29), the above numerical approximation is still valuable and $F(u)$ Eq.(B.1) can be taken in following derivations for the convenience of calculation.

\section[]{The interpolation formulation of $E(\lambda)$} \label{apd: The interpolation formulation of E_lambda}
The useful interpolation formulations about $E(\lambda)$ and its first derivative are
\begin{eqnarray}
\fl 	E(\lambda) = \frac{ \theta(1 - \lambda) }{2} ( 1 -  h_{1} )^{1/2} \int_{s}^{1} dz  \frac{ \left[ k( 1-h_{1} ) + 3  h_{1} z \right] }{ ( 1 - h_{1} + z h_{1} )^{5/2}  \left[ 1 + z^{2} ( c_{2}(1 - z^2) + c_{4}z^2 ) \right]^{1/2} } ,
\end{eqnarray}
and 
\begin{equation}
	\frac{d E(\lambda)}{ d \lambda } =  - \frac{1}{2} \frac{ k - (k-3) \lambda h_{1}
	}{ (1 - \lambda h_{1})^{1/2}  \left[ 1 + s^{2} ( c_{2}(1 - s^2) + c_{4}s^2 ) \right]^{1/2} }.
\end{equation}
Here, $s = \lambda (1 - h_{1})/(1-\lambda h_{1})$ and coefficients are
\begin{equation}
	c_{2} = -\frac{1}{4} \frac{ ( \langle h^2 \rangle - \langle h \rangle^2 ) }{ (1-\langle h \rangle)^2 },
\end{equation} 
\begin{equation}
	c_4=\frac{\left\langle(1-h)^{1 / 2}\right\rangle^2}{(1-\langle h\rangle)}-1 .
\end{equation}
The three flux-surface averaged quantities can be written as 
\begin{equation}
	\langle h \rangle = 1 - \varepsilon,
\end{equation}
\begin{equation}
	\langle h^2 \rangle = \frac{(1-\varepsilon)^2}{(1 - \varepsilon)^{1/2}},
\end{equation}
\begin{equation}
	\left\langle(1-h)^{1 / 2}\right\rangle=\frac{1}{\pi}\left[(1+\varepsilon) \sin ^{-1} \sqrt{\frac{2 \varepsilon}{1+\varepsilon}}+\sqrt{2 \varepsilon(1+\varepsilon)}\right] .
\end{equation}
By definition $E(\lambda) \equiv D_{k}(\lambda) M(\lambda)$, the interpolation form of $M(\lambda)$ is
\begin{eqnarray}
\fl 	M(\lambda)  = \frac{ \theta(1 - \lambda) }{2}  \frac{( 1 -  h_{1} )^{1/2}}{ k - (k-3) \lambda h_{1} } \cr
 \times \int_{s}^{1} dz  \frac{( k(1-h_{1}) + 3 z h_{1} )}{ ( 1 - h_{1} + z h_{1} )^{5/2}  \left[ 1 + z^{2} ( c_{2}(1 - z^2) + c_{4}z^2 ) \right]^{1/2}  } .
\end{eqnarray}
	
\section*{References}

\bibliographystyle{unsrt}
\bibliography{reference}

\begin{thebibliography}{10}

\bibitem{Fish1987_RevModPhys.59.175}
Nathaniel~J. Fisch.
\newblock Theory of current drive in plasmas.
\newblock {\em Rev. Mod. Phys.}, 59:175--234, Jan 1987.

\bibitem{Fidone198410.1063/1.864528}
I.~Fidone, G.~Giruzzi, G.~Granata, and R.~L. Meyer.
\newblock Current drive by the combined effects of electron‐cyclotron and
  landau wave damping in tokamak plasmas.
\newblock {\em The Physics of Fluids}, 27(10):2468--2476, 10 1984.

\bibitem{Fidone_1987}
I.~Fidone, G.~Giruzzi, V.~Krivenski, E.~Mazzucato, and L.F. Ziebell.
\newblock Cross-effect on electron cyclotron and lower hybrid current drive in
  tokamak plasmas.
\newblock {\em Nuclear Fusion}, 27(4):579, apr 1987.

\bibitem{SHOUCRI1989253}
M.~Shoucri, I.~Shkarofsky, V.~Fuchs, K.~Kupfer, A.~Bers, and S.~Luckhardt.
\newblock A quasilinear fokker-planck code for the numerical solution of the
  lower-hybrid current-drive problem in the presence of electron cyclotron
  heating.
\newblock {\em Computer Physics Communications}, 55(3):253--268, 1989.

\bibitem{Dumont2000combined}
R.~Dumont, G.~Giruzzi, and E.~Barbato.
\newblock {Combined kinetic and transport modeling of radiofrequency current
  drive}.
\newblock {\em Physics of Plasmas}, 7(12):4972--4982, 12 2000.

\bibitem{PhysRevLett.93.255002}
G.~Giruzzi, J.~F. Artaud, R.~J. Dumont, F.~Imbeaux, P.~Bibet, G.~Berger-By,
  F.~Bouquey, J.~Clary, C.~Darbos, A.~Ekedahl, G.~T. Hoang, M.~Lennholm,
  P.~Maget, R.~Magne, J.~L. S\'egui, A.~Bruschi, and G.~Granucci.
\newblock Synergy of electron-cyclotron and lower-hybrid current drive in
  steady-state plasmas.
\newblock {\em Phys. Rev. Lett.}, 93:255002, Dec 2004.

\bibitem{Dumont2004theory}
R.~J. Dumont and G.~Giruzzi.
\newblock {Theory of synergy between electron cyclotron and lower hybrid
  waves}.
\newblock {\em Physics of Plasmas}, 11(7):3449--3459, 07 2004.

\bibitem{Huang_22252162}
J.~Huang, E-mail:~sychen531@163.com Chen, S.~Y., C.~J. Tang, Key~Laboratory
  of~High Energy Density~Physics, and Chengdu~610064 Technology of Ministry~of
  Education, Sichuan~University.
\newblock Dependence of synergy current driven by lower hybrid wave and
  electron cyclotron wave on the frequency and parallel refractive index of
  electron cyclotron wave for tokamaks.
\newblock {\em Physics of Plasmas}, 21(1), 01 2014.

\bibitem{negative_Chen_2012}
S~Y Chen, B~B Hong, Y~Liu, W~Lu, J~Huang, C~J Tang, X~T Ding, X~J Zhang, and
  Y~J Hu.
\newblock Numerical analysis on the synergy between electron cyclotron current
  drive and lower hybrid current drive in tokamak plasmas.
\newblock {\em Plasma Physics and Controlled Fusion}, 54(11):115002, sep 2012.

\bibitem{OKCD_Zheng_2019}
P.W. Zheng, X.Y. Gong, X.H. Yin, L.H. He, X.Q. Lu, J.J. Cao, Q.H. Huang,
  S.~Deng, J.F. Lin, Y.J. Zhong, and L.~Yin.
\newblock Synergy effect of the ohkawa current drive of electron cyclotron
  waves and the lower hybrid current drive: a new mechanism.
\newblock {\em Nuclear Fusion}, 59(5):054003, apr 2019.

\bibitem{Cohen1987}
Ronald~H. Cohen.
\newblock Effect of trapped electrons on current drive.
\newblock {\em The Physics of Fluids}, 30(8):2442--2449, 08 1987.

\bibitem{LinLiu2003}
Y.~R. Lin-Liu, V.~S. Chan, and R.~Prater.
\newblock {Electron cyclotron current drive efficiency in general tokamak
  geometry}.
\newblock {\em Physics of Plasmas}, 10(10):4064--4071, 10 2003.

\bibitem{D.W.Ignat_1994}
D.W. Ignat, E.J. Valeo, and S.C. Jardin.
\newblock Dynamic modelling of lower hybrid current drive.
\newblock {\em Nuclear Fusion}, 34(6):837, jun 1994.

\bibitem{toray_matsuda1989ray}
Kyoko Matsuda.
\newblock Ray tracing study of the electron cyclotron current drive in diii-d
  using 60 ghz.
\newblock {\em IEEE transactions on plasma science}, 17(1):6--11, 1989.

\bibitem{genray_smirnov1994general}
AP~Smirnov, RW~Harvey, and K~Kupfer.
\newblock A general ray tracing code genray.
\newblock {\em Bull Amer. Phys. Soc}, 39(7):1626, 1994.

\bibitem{Antonsen198210.1063/1.863906}
Jr. Antonsen, T.~M. and K.~R. Chu.
\newblock Radio frequency current generation by waves in toroidal geometry.
\newblock {\em The Physics of Fluids}, 25(8):1295--1296, 08 1982.

\bibitem{Fisch1981_PhysRevA.24.3245}
N.~J. Fisch.
\newblock Current generation in a relativistic plasma.
\newblock {\em Phys. Rev. A}, 24:3245--3248, Dec 1981.

\bibitem{C.C.Petty_2002}
C.C. Petty, R.~Prater, J.~Lohr, T.C. Luce, W.R. Fox, R.W. Harvey, J.E. Kinsey,
  L.L. Lao, and M.A. Makowski.
\newblock Detailed measurements of the electron cyclotron current drive
  efficiency on diii-d.
\newblock {\em Nuclear Fusion}, 42(12):1366, nov 2002.

\bibitem{fc_1995}
Y.~R. Lin‐Liu and R.~L. Miller.
\newblock {Upper and lower bounds of the effective trapped particle fraction in
  general tokamak equilibria}.
\newblock {\em Physics of Plasmas}, 2(5):1666--1668, 05 1995.

\bibitem{Start_1983}
D~F~H Start, J~G Cordey, and T~Edlington.
\newblock The effect of localised wave power absorption on ecrh driven currents
  in a tokamak.
\newblock {\em Plasma Physics}, 25(4):447, apr 1983.

\bibitem{Yang2017}
Xiang~Nong Yang You-Lei, Hu Ye-Min.
\newblock Effects of trapping electrons on synergy of lower-hybrid wave and
  electron cyclotron wave.
\newblock {\em Acta Phys. Sin.}, 66(24):245202, 2017.

\bibitem{Wei_2016}
Wei Wei, Bo-Jiang Ding, Y~Peysson, J~Decker, Miao-Hui Li, Xin-Jun Zhang,
  Xiao-Jie Wang, and Lei Zhang.
\newblock Optimized calculation of the synergy conditions between electron
  cyclotron current drive and lower hybrid current drive on east*.
\newblock {\em Chinese Physics B}, 25(1):015201, dec 2015.

\bibitem{Taguchi1983}
Masayoshi Taguchi.
\newblock The effect of trapped electrons on the wave-induced current.
\newblock {\em Journal of the Physical Society of Japan}, 52(6):2035--2040,
  1983.

\bibitem{shoucri1987application}
M~Shoucri, Vladimir Fuchs, and Abraham Bers.
\newblock The application of the finite element library twodepep for the
  numerical solution of the relativistic fokker-planck equation.
\newblock {\em Computer physics communications}, 46(3):337--344, 1987.

\bibitem{Poli2018}
E.~Poli, M.~Müller, H.~Zohm, and M.~Kovari.
\newblock {Fast evaluation of the current driven by electron cyclotron waves
  for reactor studies}.
\newblock {\em Physics of Plasmas}, 25(12):122501, 12 2018.

\bibitem{Zhuang_2019}
G.~Zhuang, G.Q. Li, J.~Li, Y.X. Wan, Y.~Liu, X.L. Wang, Y.T. Song, V.~Chan,
  Q.W. Yang, B.N. Wan, X.R. Duan, P.~Fu, B.J. Xiao, and the CFETR Design~Team.
\newblock Progress of the cfetr design.
\newblock {\em Nuclear Fusion}, 59(11):112010, jun 2019.

\bibitem{Prater_2008}
R.~Prater, D.~Farina, Yu. Gribov, R.W. Harvey, A.K. Ram, Y.-R. Lin-Liu,
  E.~Poli, A.P. Smirnov, F.~Volpe, E.~Westerhof, A.~Zvonkov, and the ITPA
  Steady State Operation Topical~Group.
\newblock Benchmarking of codes for electron cyclotron heating and electron
  cyclotron current drive under iter conditions.
\newblock {\em Nuclear Fusion}, 48(3):035006, jan 2008.

\end{thebibliography}

\end{document}